\documentclass[sigplan,10pt,manuscript,nonacm]{acmart}
\acmISBN{} 
\acmDOI{} 
\startPage{1}

\setcopyright{none}

\usepackage[parfill]{parskip}

\PassOptionsToPackage{table}{xcolor}

\usepackage{colortbl}
\usepackage{amsmath}

\usepackage{color}
\usepackage{amssymb}
\usepackage{algorithmic}
\usepackage{endnotes}
\usepackage{epsfig}
\usepackage{epstopdf}
\usepackage{caption}
\usepackage{subcaption}
\usepackage{url}
\usepackage{svg}
\usepackage{color}
\colorlet{shadecolor}{gray!30}
\usepackage{graphicx}
\usepackage[utf8]{inputenc}

\usepackage[breakable]{tcolorbox}

\usepackage[frozencache,cachedir=.]{minted}
\definecolor{CodeBG}{gray}{0.9}

\usepackage{pgfplots,tikz}
\usepackage{tikz-network}
\usetikzlibrary{arrows,automata}
\usetikzlibrary{external}
\pgfplotsset{compat=1.16}
\usepackage{wasysym}
\usepackage{listings}
\usepackage{bm}
\usepackage{relsize}
\usepackage{paralist}
\usepackage{url}
\usepackage{comment}
\usepackage{mdframed}
\usepackage{booktabs}
\usepackage{wrapfig}
\usepackage[toc,page]{appendix}
\usepackage{wasysym}
\usepackage{listings}
\usepackage{setspace}
\usepackage{float}
\usepackage{longtable}
\usepackage[nodayofweek]{datetime}
\usepackage{setspace}
\usepackage{tabularx}
\usepackage[inline,shortlabels]{enumitem}

\definecolor{lbcolor}{rgb}{0.9,0.9,0.9} 
\usetikzlibrary{backgrounds}
\usetikzlibrary{patterns}

\newcommand*\wrapletters[1]{\wr@pletters#1\@nil}
\def\wr@pletters#1#2\@nil{#1\allowbreak\if&#2&\else\wr@pletters#2\@nil\fi}

\newcommand{\code}[1]{\sloppy{\texttt{#1}}}

\usepackage{xargs}

\newcommandx{\changethis}[2][1=]{\todo[linecolor=red,backgroundcolor=red!25,bordercolor=red,#1]{#2}}
\newcommandx{\changefixed}[2][1=]{\todo[linecolor=green,backgroundcolor=green!25,bordercolor=green,#1]{#2}}
\newcommandx{\unsure}[2][1=]{\todo[linecolor=purple,backgroundcolor=purple!25,bordercolor=purple,#1]{#2}}
\newcommandx{\thiswillnotshow}[2][1=]{\todo[disable,#1]{#2}}
\usepackage{ amssymb }

\newcommand{\academic}{$^\mathscr{A}$}
\newcommand{\technical}{$^\mathcal{T}$}

\begin{document}

\title{MetaFFI - Multilingual Indirect Interoperability System}
\titlenote{Supported by Len Blavatnik and the Blavatnik Family foundation}

\author{Tsvi Cherny-Shahar}
\affiliation{
  \department{Blavatnik School of Computer Science} 
  \institution{Tel Aviv University}                 
  \country{Israel}
}
\email{tsvic@mail.tau.ac.il}                        
\author{Amiram Yehudai}
\affiliation{
  \department{Blavatnik School of Computer Science} 
  \institution{Tel Aviv University}                 
  \country{Israel}
}
\email{amiramy@tau.ac.il}

\begin{abstract}
The development of software applications using multiple programming languages has increased in recent years, as it allows the selection of the most suitable language and runtime for each component of the system and the integration of third-party libraries. However, this practice involves complexity and error proneness, due to the absence of an adequate system for the interoperability of multiple programming languages. Developers are compelled to resort to workarounds, such as library reimplementation or language-specific wrappers, which are often dependent on C as the common denominator for interoperability. These challenges render the use of multiple programming languages a burdensome and demanding task that necessitates highly skilled developers for implementation, debugging, and maintenance, and raise doubts about the benefits of interoperability. To overcome these challenges, we propose MetaFFI, a pluggable in-process indirect-interoperability system that allows the loading and utilization of entities from multiple programming languages. This is achieved by exploiting the less restrictive shallow binding mechanisms (e.g., Foreign Function Interface) to offer deep binding features (e.g., object creation, methods, fields). MetaFFI provides a runtime-independent framework to load and \emph{xcall} (Cross-Call) foreign entities (e.g., functions, objects). MetaFFI uses Common Data Types (CDTs) to pass parameters and return values, including objects and complex types, and even cross-language callbacks. The indirect interoperability approach of MetaFFI has the significant advantage of requiring only $2n$ mechanisms to support $n$ languages, as opposed to the direct interoperability approaches that need $n^2$ mechanisms. We have successfully tested the binding between Go, Python3.11, and Java in a proof-of-concept on Windows and Ubuntu.
\end{abstract}


\maketitle

\section{Terminology}
This article uses the following terminology.
\begin{itemize}
    \vspace{-2mm}\item Language - a pair of (Syntax,Runtime)
    \vspace{-2mm}\item Programming Language - a Language used to write the logic layer of a program
    \vspace{-2mm}\item Host Language (or Host) - Programming language initiating a call to a different programming language.
    \vspace{-2mm}\item Guest Language (or Guest) - Programming language implementing the called code
    \vspace{-2mm}\item Foreign Entity – Function, class, method, field (etc.) in the guest language
    \vspace{-2mm}\item Shallow Interoperability Binding – Basic accessibility to guest Language (e.g. calling function)
    \vspace{-2mm}\item Deep Interoperability Binding – Broad accessibility to guest language (access to objects/types)
    \vspace{-2mm}\item Foreign Function Interface (FFI) \cite{wiki_ffi} – A shallow binding mechanism that provides the ability to call a function from a single Host to a single Guest (one way)
    \begin{itemize}
        \vspace{-2mm}\item C-FFI - An FFI mechanism binding calling C from host language
    \end{itemize}
    \vspace{-2mm}\item Interoperability mechanisms - Mechanisms allowing to use multiple languages in the same operating system process (e.g. FFI, runtime embedding)
    \vspace{-2mm}\item Language Port - A new programming language with the same syntax as the porting programming language but using a different runtime environment (e.g. Jython\cite{jython}, IronPython\cite{ironpython}, JRuby\cite{jruby})
\end{itemize}

\section{Introduction}\label{sec:intro}
Interoperability between languages is a long-standing challenge that has become more acute in recent years, as applications require more functionality and performance, requiring more complex software and hardware \cite{challenges_of_interop}\cite{empirical_multi_lingual}. Therefore, it is improbable that a single programming language or runtime can adequately serve an entire application, and in some cases confining an entire application to one programming language is infeasible. Since the emergence of the first programming languages in the late 1950s, hundreds of new languages have been developed, some for general purposes and others for specific domains. Despite the availability of a wide variety of languages and runtimes for different tasks, it is difficult to use more than one programming language within a single application \cite{empirical_multi_lingual}, as languages typically do not support interoperability between them or only with a limited number of languages.

Moreover, the reluctance to use multiple programming languages results in the reimplementation and rewriting of existing code from one language to another (e.g., porting), rather than reusing the existing code, even if the runtime of the original library offers better performance, maintainability, and popularity \cite{empirical_multi_lingual}. 

Although the benefits of reusing existing code are evident, reusing code between multiple languages in a single application is a complex and error-prone task in itself, as this kind of development obliges developers to devote considerable effort to cross-language techniques that are difficult to use and maintain \cite{empirical_multi_lingual}\cite{xlang_survey}.

Several solutions have been proposed to address this problem: \begin{itemize} \item Foreign Function Interface (FFI) \cite{wiki_ffi} enables invocation of cross-language functions, which we call \emph{shallow binding}, as it allows access to functions, but not to objects, fields, globals, etc. We refer to entities implemented in a different language as \emph{foreign entities}. \item Virtual machines such as JVM (Java Virtual Machine) \cite{jvm_specs} and Microsoft CLR (Common Language Runtime) \cite{crl_overview}, facilitating direct access to foreign entities, which we call \emph{deep binding}, but only for languages compiled to the virtual machine’s \textit{executable code} and running within the virtual machine runtime. \item Language-independent object models such as CORBA (Common Object Request Broker Architecture) \cite{corba} and COM (Common Object Model) \cite{com_component}, which provide access to objects in multiple languages by specifying a well-known structure of the interoperable objects. This approach is restrictive and might require from a programming language features it doesn't necessarily support.\end{itemize}

Most existing solutions for language interoperability focus on binding a pair of languages (e.g., Python ctypes \cite{python_ctypes} as C-interface FFI) or a small set of languages (e.g., VM languages that compile to the same VM bytecode), but primarily interoperate with C \cite{empirical_multi_lingual}. Furthermore, other solutions are challenging and cumbersome to use \cite{empirical_multi_lingual}. Chisnall \cite{challenges_of_interop} highlighted the significance and difficulties of language interoperability in 2013. As the number of languages grows, encompassing both general-purpose and domain-specific languages, the selection and ability to select the appropriate tool for each task become more crucial and more complicated. 

To address the challenges mentioned above, we define simple interoperability, a collection of four essential features that we believe are needed to facilitate interoperability:
\begin{enumerate}
\item Host-Only Coding - Use guest entities without the need of understanding or utilizing the guest language
\item No manual IDL - User is not required to write a dedicated IDL for interoperability
\item Automatic runtime management - User is not required to hassle with the different runtimes options and internals (e.g., classpath \cite{classpath} when using JVM)
\item Mappable Common Data Type - the system provides set of data types that can be used by all languages, and can be mapped to data types in all supported languages
\end{enumerate}
We propose MetaFFI, a pluggable in-process indirect-interoperability system that support the simple interoperability as defined above. The choice of these four features is explained and detailed in the empirical study on the multilingual development and interoperability of programming languages \cite{empirical_multi_lingual}. The empirical study shows the popularity of multi-PL among open-source projects and interoperability tools. It discusses what is needed for the multi-PL system and defines \emph{simple interoperability}, which is the set of features presented above.

MetaFFI interoperability concept derived from the exported entities loading mechanisms of dynamic libraries provided by popular operating systems (e.g., LoadLibrary\cite{loadlibrary_msdn}/dlopen\cite{dlopen_die} and GetProcAddress\cite{getprocaddress_msdn}/dlsym\cite{dlsym_die}). In the same manner an application can load a binary, platform-specific, dynamic library and import its exported entities, MetaFFI provides a layer to load modules and their entities in multiple formats while providing a uniform API.

MetaFFI enables deep binding capabilities using only shallow binding mechanisms, facilitating the development of independent modules irrespective of the language of implementation. To use MetaFFI, the user performs the following steps:
\begin{itemize}
\item loading the runtime using the \code{load\_runtime} function
\item loading the module using the \code{load\_module} function, which is analogous to LoadLibrary \cite{loadlibrary_msdn} in Windows or dlopen \cite{dlopen_die} in Linux
\item loading the required foreign entities using the \code{load\_entity} function, which analogous to GetProcAddress \cite{getprocaddress_msdn} in Windows or dlsym \cite{dlsym_die} in Linux 
\end{itemize}

MetaFFI adopts an indirect approach to link languages via a common C-interfaced module, rather than directly connecting them. This reduces the number of FFI and interoperability libraries needed to support $n$ languages in a project from $O(n^{2})$ to $O(n)$. In addition, MetaFFI’s modular plugin design allows developers to extend the programming language support by implementing up to three C-interfaces, which facilitate the interoperability between the newly added programming language and all the existing supported languages. The complexity of adding a new language does not depend on the number of languages already supported. Plugins are integrated by implementing their C interface. The details of the plugin design are discussed in section \ref{sec:metaffi_system}.

To evaluate the effectiveness of our framework design, we have implemented a fully functional prototype. We have carried out experiments on the system using \emph{Go}, \emph{Python3} and \emph{OpenJDK Java}, on Windows 11 and Ubuntu 22.04 platforms. The test suite covers both user-defined source code and third-party libraries.

The rest of the paper is organized as follows: Section \ref{sec:backgroud} discusses previous and related work with respect to our research. Section \ref{sec:usage_example} presents MetaFFI with a usage example and explains the MetaFFI API. Section \ref{sec:metaffi_system} details the MetaFFI system, including its design, mechanisms, components, common data type, and limitations.  Section \ref{sec:add_lang} describes in further detail the process of adding a new language support. Section \ref{sec:idl-details} details the MetaFFI IDL structure and the process of adding a new IDL generator.  Section \ref{sec:future_work} presents future work and research in the MetaFFI system. Finally, Section \ref{sec:conclusions} concludes the paper.

Sections addressing an academic audience are marked with \academic, while sections addressing a technical audience are marked with \technical . Unmarked sections address all readers.

\section[Background and Related Work]{Background and Related Work \academic}
\label{sec:backgroud} 

SWIG (Simplified Wrapper and Interface Generator) \cite{swig}, which was started by Beazley in 1996, is a tool that generates glue code to access \code{C}/\code{C++} code from 19 languages (as of this writing). To access the exported code in \code{C} or \code{C++}, the developer must manually declare the target function using SWIG’s Intermediate Definition Language (IDL\cite{idl_wikipedia}). The SWIG tool then produces the glue code in the desired language to invoke the exported function. Thus, SWIG is a \emph{compile-time} solution to call binary-compilated functions (also known as \emph{native functions}). SWIG primarily supports binding with C and C++, and has limited support for bidirectional language interoperability. It also requires the developer to have knowledge of C/C++ and SWIG IDL. It poses a high learning challenge in complex scenarios \cite{swig_master_class}. MetaFFI, on the other hand, does not rely on an IDL file and can support a wide range of languages, not just C/C++. Furthermore, MetaFFI offers bidirectional support, including passing callbacks to guest languages.

\emph{LibFFI} project \cite{libffi}, which was initiated in 1996 by Green, offers the capability to invoke native (i.e. binary) compiled functions at runtime based on the information provided about the functions. LibFFI is a runtime solution, in contrast to the compile-time solution that SWIG provides. This feature enables LibFFI users to call C functions with unknown signatures at compile time. For instance, Python \code{ctypes} \cite{python_ctypes} (FFI from Python to \code{C}) utilizes LibFFI to invoke native compiled functions. It shares the same objective of calling native functions with SWIG, but differs from SWIG in that it generates the appropriate function call at runtime. Like SWIG, LibFFI supports native functions and not multiple languages. MetaFFI, on the other hand, adopts a similar concept to LibFFI but aims to support not only native functions, but also multiple entities (e.g. methods, fields) in different languages and runtimes. MetaFFI may use LibFFI as shallow binding to provide a deep binding capabilities, like in the case of binding Python3 using CTypes \cite{python_ctypes}, which uses LibFFI.

GLib Object System (GObject) \cite{gobject} is a library that implements an object system for C. Using GObject, GObject Introspection (GIR) \cite{gir} provides interoperability between C libraries and language bindings. In addition to the fact that GObject only supports interaction with C, it has a steep learning curve, effort, and requires some boilerplate code \cite{gobject_wiki}\cite{gobject_boilerplate}. In addition, GIR cannot support $3^{rd}$ party libraries out of the box without wrapping them in GObject. Vala \cite{vala}, a compiler from C\# to C, supports GObjects natively, but it supports only single direction interoperability (C\#$\rightarrow$C) and adding support to more bindings requires a great effort. 

Haxe \cite{haxe} is a high-level programming language that compiles to several other programming languages (e.g., C++, JavaScript, and more) and can be executed directly on a HashLink virtual machine. Although Haxe supports compiling the syntax of several languages, it does not execute the guest language in its original runtime. Thus, it does not support guest languages under our definition as a pair of (syntax, runtime). Therefore, Haxe provides languages ports that reimplement the syntax on a different runtime (e.g., Jython \cite{jython}, IronPython \cite{ironpython}, and \cite{jruby}). Moreover, the single-language approach prevents developers from choosing the language they actually wish to use (both syntax and runtime).

The LLVM \cite{llvm} project, started in 2000 by Adve et al., specifically the LLVM intermediate representation (LLVM-IR), allows compilers to compile code in a given language (syntax) to LLVM-IR, a common low-level language, which can be later compiled to binary. By compiling several languages into LLVM-IR, code in one language can call functions in other languages to achieve interoperability. On the one hand, using LLVM achieves high performance and low overhead for interoperating between languages. On the other hand, it is "losing" the original runtime of the language, resulting in a language port. Thus, it does not support interoperability between many languages. An LLVM language port requires a tremendous amount of work -- writing a new front-end compiler to LLVM-IR (if such a front-end does not exist), which needs to implement the different runtime features (for example, Goroutines in Go). This approach ignores the fact that these features have already been implemented and tested successfully in the original programming language runtime. In addition, future runtime updates (like JVM or .Net updates with improvements) would also need to be implemented in the compiler, resulting in two implementations of the same logic. Other interesting and notable solutions have taken various paths to provide cross-language programming for a large number of languages without FFI. The virtual machine provides \emph{deep binding} between all the languages running on that virtual machine. For example, Microsoft .NET framework \cite{ms_dotnet} which executes Common Intermediate Language (CIL, formerly known as MSIL) \cite{ms_cil} using the JIT compiler.

TruffleVM \cite{trufflevm} by Grimmer et al. is a framework that proposes a different approach to achieve "deep binding" between multiple languages by using JVM and creating JVM versions for non-JVM languages. TruffleVM is a virtual machine that supports "Truffle-based" languages such as TruffleC, TruffleRuby, and TruffleJS, which are JVM-based versions of the languages, running on the HotSpot JVM engine \cite{jvm_hotspot}. The authors' goal is to enable full interoperability between these languages, similar to what Microsoft has accomplished with their CLR languages (\code{C\#}, \code{C++.Net}, \code{VB.Net} and others). This approach offers a method to create language ports, but it faces the same challenges that language ports entail (such as different behavior, version differences, huge effort, bug fixing, maintenance, etc.). It also confines the user to the virtual machine. The VM approach fails to provide interoperability between multiple languages outside the virtual machine, whereas MetaFFI strives to maintain the language runtime, thus enabling interoperability to many languages that TruffleVM cannot support.

A paper by Kaplan et al. \cite{polispin} criticizes the usage of IDL for imposing potentially a serious overhead on software developers, and presents an IDL-invisible approach using PolySPIN, which tries to automatically map data types from a specific host language to a specific guest language using a \emph{matcher} and generates code which maps these types automatically, removing the need for IDL. Although we agree that manually writing an IDL is an overhead (whether in a dedicated syntax as in SWIG or using the host language syntax as in Python CTypes \cite{python_ctypes}), PolySPIN requires extensive, very language-specific implementations that parse and deeply understand the supported languages by matching the type systems. Although the approach is interesting, supporting $n$ languages would require $n^2$ matchers. Moreover, the effort to implement a \emph{matcher} from one language to another produces a vast overhead in adding language support to a large number of languages. We do agree that requiring a manually crafted IDL is a tedious task for the developer, especially for large libraries where it becomes impractical.

A paper by Wegiel et al. \cite{colors} presents a Co-Located Runtime Sharing (CoLoRS) which provides a transparent shared memory across languages and across runs with a language-neutral object/class model in a CoLoRS server process that manages the shared memory of other processes, making the objects language independent. This approach can be an alternative to our approach of using CDTs (section \ref{sec:cdt}) to create a common data type for both host and guest languages, but in our opinion an out-of-process solution like CoLoRS is more error prone and needlessly complex, introducing challenges for cleanup, failed processes, and more.

The following papers present interesting approaches but provide only interoperability with C. A paper by Turcotte et al. \cite{pos_lua} shows a technique to perform FFI without modeling the guest language. The paper presents an FFI from \code{Lua} to \code{C} using a Lua-typed POC called \emph{Poseidon Lua}. Poseidon Lua provides allocation and accessibility to \code{C}-data enabling us to call \code{C}-functions, all by using \code{Poseidon Lua} types. Direct access to foreign language memory raises issues of mutability, as discussed in \cite{challenges_of_interop}. The paper provides interoperability with C and does not meet the definition of \emph{simple interoperability}. Another paper by Yallop et al. \cite{generic_interop} shows a method of using generic programming to define and build declarative functions of the guest language in the host language. The authors show a binding from \code{OCaml} to \code{C} based on the generic programming features available in \code{OCaml} to implement the \code{C}-Types in \code{OCaml}. The idea of using generic programming is interesting, but requires the host language to support generic programming. Moreover, the technique suffers from mutability issues raised in \cite{challenges_of_interop} due to its direct access to foreign values.

The papers discussed in this section can be leveraged by MetaFFI to support more languages, but not all interoperability approaches are compatible with MetaFFI. For instance, a paper by Chiba et al. \cite{code_migration} proposes a framework for executing a "code block" in a foreign language by enhancing the "string embedding" technique, which involves writing the foreign code as a string and running it using the \code{eval} function of the foreign language. This technique is only suitable for host and guest languages that offer \code{eval} functionality.

\section{Usage Example} \label{sec:usage_example}

MetaFFI provides the functionalities mentioned in the Introduction (i.e. \code{load\_runtime} and \code{load\_entity}), in the Cross-Language Link Runtime (XLLR) (section \ref{sec:xllr}) through C-API. Each supported language also provides a MetaFFI API that wraps the XLLR-API and abstracts the usage of C-FFI from the end-user, making the usage simpler by not requiring any C-FFI knowledge from the MetaFFI user.

The \code{load\_entity} function returns an XCall structure that contains a C function pointer to the foreign entity entrypoint, and a context structure containing call-related data. In the event that an entity is a data member or attribute, the returned C function acts as a getter or setter. The parameters and return values are passed using Common Data Types (CDTs) detailed in section \ref{sec:cdt}.  MetaFFI APIs also wrap the returned XCall structure to simplify usage.

The manual loading of entities works well when the number of entities is small, but in the case of large libraries, writing the code to load each entity is a hefty task. To mitigate this problem, MetaFFI supports a host compiler (section \ref{sec:host-compiler}) that generates the runtime and foreign entities loading code and creates wrapper entities in the host language that wrap the foreign entities.

\begin{listing}
\begin{singlespace}
\begin{minted}[bgcolor=CodeBG,fontsize=\footnotesize,autogobble,tabsize=2,linenos]{python}
# load JVM
runtime = MetaFFIRuntime('openjdk')

# load log4j
log4j_api_module = runtime.load_module('log4j-api-2.21.1.jar;log4j-core-2.21.1.jar')

# load getLogger() method to get a new logger
getLogger = log4j_api_module.load('class=org.apache.logging.log4j.LogManager,callable=getLogger', 
    [metaffi_string8_type], 
    [new_metaffi_type_with_alias(metaffi_handle_type, 'org.apache.logging.log4j.Logger')])

# load error() method in logger
perror = log4j_api_module.load('class=org.apache.logging.log4j.Logger,callable=error,instance_required', 
    [new_metaffi_type_with_alias(MetaFFITypes.metaffi_handle_type), 
    new_metaffi_type_with_alias(MetaFFITypes.metaffi_string8_type)],  
    None)

# create logger with getLogger()
logger = getLogger('pylogger')
perror(logger, 'Logging error from python!')
\end{minted}
\end{singlespace}
\caption{Log4J from Python3.11}
\label{code:log4j-from-py3}
\end{listing}

Listing \ref{code:log4j-from-py3} demonstrates how Python3.11 can use Log4J \cite{log4j2}, a popular Java log library that has many ports to other languages (\cite{log4net}\cite{log4cxx}\cite{log4python} and more). The code uses the Python3.11 MetaFFI API, which wraps calls to the XLLR C-API. \begin{itemize}
\item Line 2 loads the JVM runtime
\item Line 5 loads the modules required for Log4J
\item Line 8 loads the \code{LogManager.getLogger()} static method, which returns a log4j logger instance. \code{load} expects a \textit{function path} (section \ref{sec:function-path}) and MetaFFI Types (section \ref{sec:metaffi-type}) used as parameters and return values.
\item Lines 13-16 loads the \code{Logger.error()} method, which prints an error using Log4J
\item Line 19 creates a new logger named \textit{pylogger}
\item Line 20 calls the \code{logger.error()} method.
\end{itemize}

Instead of manually loading foreign entities, the user can execute in the terminal:\\\code{metaffi -c --idl log4j-api-2.21.1.jar -h}, which invokes the MetaFFI host compiler, generating Python entities that wrap log4j entities, creating a smoother user experience without writing code that interacts with the MetaFFI API.

The underlying mechanism of MetaFFI makes foreign entities available as C functions, these function pointers can be passed to a foreign entity as arguments. Therefore, MetaFFI also provides pass callback functions as shown in listing \ref{code:py3-callable}.

\begin{listing}[t]
\begin{singlespace}
\begin{minted}[bgcolor=CodeBG,fontsize=\footnotesize,autogobble,tabsize=2,linenos]{python}
# runtime_test_target.py
def call_callback_binary_op(bi_op_func: Callable) -> int:
	res = bi_op_func(1, 2) # call binary function
	return res # return result
\end{minted}
\begin{minted}
[bgcolor=CodeBG,fontsize=\footnotesize,autogobble,tabsize=2,linenos]{java}
public static int add(int x, int y)  // java add
{
		return x+y;
}
\end{minted}
\begin{minted}
[bgcolor=CodeBG,fontsize=\footnotesize,autogobble,tabsize=2,linenos]{java}
// load python3.11 runtime
var runtime = new MetaFFIRuntime("python311");
runtime.loadRuntimePlugin();

// load runtime_test_target.py
var module = runtime.loadModule("./python3/runtime_test_target.py");

// load the python "call_callback_binary_op" function
metaffi.Caller callCallback = module.load("callable=call_callback_add",
    new MetaFFITypeInfo[]{ new MetaFFITypeInfo(MetaFFITypes.MetaFFICallable) },
    null);

// must OpenJDK runtime to handle callbacks
var javaRuntime = new MetaFFIRuntime("openjdk");
javaRuntime.loadRuntimePlugin();

// use reflection to get the "add" method
Method m = TestClass.class.getDeclaredMethod("add", int.class, int.class);

// make the method callable from MetaFFI
metaffi.Caller callbackAdd = api.MetaFFIRuntime.makeMetaFFICallable(m);

// call python passing the "add" function has parameter
callCallback.call(callbackAdd);
\end{minted}
\end{singlespace}
\caption{Java callback method passed and called from Python3.11 function}
\label{code:py3-callable}
\end{listing}

The Python3.11 function \code{call\_callback\_binary\_op}, defined in Listing \ref{code:py3-callable}, takes as an argument a function that operates on two \code{int} values and returns an \code{int} value. The Java static method \code{add} sums the two integers and returns the result. In the final code segment, Java employs the MetaFFI JVM API to pass the \code{add} method to the \code{call\_callback\_binary\_op} function:
\begin{itemize}
    \item lines 2,3 - loads Python3.11 runtime
    \item lines 6 - loads \code{runtime\_test\_target.py} module
    \item line 9 - loads \code{call\_callback\_add} function
    \item line 14,15 - loads MetaFFI JVM runtime to handle callbacks
     \item line 18 - load "add" method
     \item line 21 - Wrap Java "add" method and make it MetaFFI enabled
     \item line 24 - Call Python3.11 \code{call\_callback\_binary\_op}, passing Java \code{add} method 
\end{itemize}

MetaFFI embeds multiple runtimes in the same process, but in many cases these runtimes are isolated and have no direct access to other runtimes. However, the C runtime is shared among all runtimes and serves as a bridge between them. The CDT structure and the communication between the runtimes uses mixed shared memory and message passing model inspired by Microsoft Object Linking and Embedding (OLE) model \cite{ole} and the Remote Procedure Call (RPC) model proposed by Nelson \cite{rpc}. If the runtimes are isolated, a message passing approach is employed, whereas if the runtimes can share memory, a shared memory approach is utilized. Details of runtime management and the invocation of cross-language functions are presented in sections \ref{sec:runtime} and \ref{sec:xcall}, respectively.

\section{MetaFFI System}
\label{sec:metaffi_system}
MetaFFI is a flexible and adaptable system that aims to indirectly facilitate multilingual compatibility. It is structured to fulfill the "simple interoperability" criteria as outlined in section \ref{sec:intro}, and presented in more detail in [9]. Through MetaFFI, external components can be integrated and utilized within the primary language's source code. The system is designed in an indirect manner, ensuring that the incorporation of new languages is not constrained by the existing language base. In addition, the inclusion of a new programming language automatically ensures compatibility with all previously integrated languages. Nevertheless, it is essential for the runtime-executed code to have the capability to call C functions for accessing other languages, as well as to be accessible from C to enable full-duplex cross-language invocation.

MetaFFI offers a C-Interface API for multilingual communication: \begin{itemize}
    \item  \code{load\_runtime\_plugin} and \code{free\_runtime\_plugin} - These functions load or unload a runtime (such as JVM, CPython, etc.).
    \item \code{load\_entity} and \code{free\_function} - These functions load a foreign entity from a specified module and return it as a C/XCall\footnotemark  function and context as \code{void**}. \code{free\_function} releases the function and context.
    \item \code{make\_callable} - This function wraps a callable in a XCall structure.
\end{itemize}

\footnotetext{XCall is an extension of C function, hence XCall function pointer is a valid C function pointer.}

The language support implementer should provide an API in the supported language, which wraps MetaFFI’s C-API, to enhance the user experience. Currently, the system offers an API for users in Go, JVM, and Python3.11.

As discussed in section \ref{sec:intro} and shown in the code listings \ref{code:py3-callable}, MetaFFI has four steps:
\begin{itemize}
    \item Loading the runtime (section \ref{sec:xllr})
    \item Loading the module
    \item Loading the foreign entity (section \ref{sec:load-entity})
    \item Calling the foreign entity (section \ref{sec:xcall})
\end{itemize}

The code listing \ref{code:py3-callable} also shows the usage of cross-language callback.

\subsection{XLLR, Runtime Plugins \& Runtime Management} \label{sec:xllr}\label{sec:runtime}
Cross-Language Link Runtime (XLLR) is the runtime component that handles the runtimes of the different languages and offers a C-API to load foreign entities (using MetaFFI plugins) that return XCall functions. 

The XLLR C-API for runtime management consists of:
\begin{minted}[bgcolor=CodeBG,fontsize=\footnotesize,autogobble,tabsize=2,linenos]{c}
void load_runtime_plugin(const char* runtime_plugin_name, char** err);
void free_runtime_plugin(const char* runtime_plugin, char** err);
\end{minted}

The function \code{load\_runtime\_plugin} takes \code{runtime\_plugin\_name} as an argument, which is the \textit{runtime plugin} to be loaded. XLLR loads the runtime plugin and forwards the call to the runtime plugin that provides the same API (excluding the \code{runtime\_plugin\_name} parameter).

The runtime plugins load (or free) their respective runtimes. For example, Python3.11’s plugin load (or attach) the Python3.11 interpreter, while JVM’s plugin create (or attach) the JVM. In case of an error, it returns to the caller via \code{err} as a heap-allocated null-terminated string.

The \code{load\_runtime\_plugin} and \code{free\_runtime\_plugin} functions implemented by the plugin must account for multiple invocations. XLLR ensures that concurrent calls to these functions are synchronized. If an error is returned, the caller is responsible for releasing \code{err}.


\subsubsection{Memory Allocation \& Deallocation} \label{sec:mem_alloc_dealloc}
In a multilingual environment, it is common for different components to be built with various compilers. This is also true for C-FFI of different runtimes. While many programming languages can interoperate with C due to using the same Application Binary Interface (ABI), this does not imply that the toolsets (such as Visual C++, Clang, TDM, etc.) used to handle C are identical. Different toolsets may implement functions like \code{malloc} and \code{free} in different ways, which can lead to application crashes or undefined behavior if malloc and free from different toolsets are mixed. To address the variety of possible toolsets, any memory allocated and returned to the caller, such as error strings or heap return values, must be allocated using the \code{xllr\_malloc} and \code{xllr\_free} functions provided by XLLR. This ensures that the same toolset is used for both allocation and deallocation. 

\subsection{Loading Foreign Entity}\label{sec:load-entity}

The XLLR C-API provides the following function to load foreign entity and return its XCall structure:
\begin{minted}[bgcolor=CodeBG,fontsize=\footnotesize,autogobble,tabsize=2,linenos]{c}
struct XCall{	void* pxcall_and_context[2]; };
struct XCall* load_function(const char* runtime_plugin_name,
        const char* module_path,
        const char* function_path,
        metaffi_type_info* params_types, int8_t params_count
        metaffi_type_info* retval_types, int8_t retval_count,
        char** err);
\end{minted}

The call is forwarded to the runtime plugin (excluding the \code{runtime\_plugin\_name} argument).

The function expects the following arguments :
\begin{itemize}
    \item \code{runtime\_plugin\_name} which specifies the runtime plugin to use.  
    \item \code{module\_path} specifies the module the entity relies in (detailed in \ref{sec:module-path}).
    \item \code{function\_path} is a string declaring the location of the entity within the module (detailed in \ref{sec:function-path}).
    \item \code{params\_types} and \code{retvals\_types} are an arrays of \code{metaffi\_type\_info}, specifying the types the entity expects and returns (detailed in \ref{sec:metaffi-type}).
    \item \code{params\_count} and \code{retval\_count} are the sizes of the array params\_types and retval\_types.
    \item \code{err} is an \texttt{out} parameter (that is, the parameter is set by the called function) of a null-terminated string. if \code{err} $\neq$ \code{NULL}, then the function failed. The caller is required to free the error string using \code{xllr\_free}.
\end{itemize}

\subsubsection{Module Path} \label{sec:module-path}

A module path refers to the module (or modules) managed by the specified runtime. The runtime plugin documentation should specify the required module path format. For instance, in JVM, the module path consists of a list of \code{.jar} files and/or directories containing the entity's dependencies. In Python3.11, the module path can be a directory containing the Python module, a path to a \code{py} file, or a package name.

\subsubsection{Function Path} \label{sec:function-path}

Windows \code{GetProcAddress} and Linux \code{dlsym} receive a string or ordinal to locate the exported function in the export table. C++ also uses the export table using name mangling \cite{name-mangling}  to "encode" additional information in the exported function name.

MetaFFI adopts the same basic idea, but an export table is not available in many programming language modules. Instead, MetaFFI uses the string to let the plugin know how to locate the entity using a simple list of attributes written in key-value pairs and tags encoded in a human readable text separated by a comma: \code{[$Key_1=Val_1,Tag_1...Tag_N,Key_N=Val_N$]}. The runtime plugin documentation must document which keys and tags it requires. Although we strive to keep the keys uniform among the plugins, it is not mandatory.

For instance, the key \emph{callable}, which is common to all the existing plugins, denotes a name of a callable entity, such as a function, constructor, or a method. Another shared tag is \emph{instance\_required}, which signifies that the entity requires an instance and that the first parameter is that instance. The \emph{global} key is used in Go to declare that a variable is in global scope, something that is not used in the JVM plugin. Likewise, the key \emph{attribute} refers to an attribute in Python3.11 and is not used in the Go or JVM plugin.

Each runtime plugin specifies the key-value pairs and tags it expects when loading an entity. The function path is then passed to the runtime plugin, which uses this information to load the foreign entity and return an XCall struct, which contains a C function pointer to the foreign entity.

\subsection{MetaFFI Enabled Callable }

The function \code{load\_entity} takes the module path and the function path as input and returns a pointer to an XCall structure, which is used to invoke the entity. The function \code{make\_callable} performs a similar task, but does not require the entity to be located in a module. Instead, it requires a callable entity within the runtime. \code{make\_callable} in XLLR forwards the call to the plugin supporting the intended runtime, which knows how to handle a callable entity in its supported runtime, which in turn creates an XCall to call that entity. The function is defined as follows:

\begin{minted}[bgcolor=CodeBG,fontsize=\footnotesize,autogobble,tabsize=2,linenos]{c}
struct XCall* make_callable(const char* runtime_plugin_name, 
                void* pcallable_entity,
                metaffi_type_info* params_types, int8_t params_count,
                metaffi_type_info* retval_types, int8_t retval_count,
                char** err);
\end{minted}

The call is forwarded to the matching runtime plugin that exports the same signature (excluding \code{runtime\_plugin\_name}).

The parameter \emph{pcallable} represents a callable entity in the target runtime. For instance, in Python3.11 it is \code{PyObject*} that refers to a callable object. In JVM it is a \code{jobject} that denotes a method.

\subsection{Common Data Types} \label{sec:cdt}

In the general case, MetaFFI assumes that all runtimes are isolated with a common shared memory, the XLLR and its plugins. In order to construct a message containing all the arguments and return values, MetaFFI uses an array of Common Data Type (CDT) structures inspired by Microsoft's OLE VARIANT structure \cite{ole}\cite{com_variant} aimed at solving a similar goal of passing a message between processes on the same computer. CDT, as opposed to VARIANT, is designed specifically for the interoperability of different programming languages. Each CDT stores a MetaFFI data type. The different types are detailed in section \ref{sec:metaffi-type}.

\begin{listing}[h]
\begin{singlespace}
\begin{minted}[bgcolor=CodeBG,fontsize=\footnotesize,autogobble,tabsize=2,linenos]{c}
struct cdt
{
	metaffi_type type; // The MetaFFI Type of the CDT - uint64
	metaffi_bool free_required; // cdt_val's data need to be freed - uint8 
	union cdt_types cdt_val; // Union storing the relevant MetaFFI data type
};
\end{minted}
\end{singlespace}
\caption{Common Data Type}
\label{code:cdt}
\end{listing}

The code listing \ref{code:cdt} defines the structure \code{cdt}. The \code{cdt.type} field specifies the MetaFFI type of the CDT, which is an unsigned 64bit integer.  The \code{cdt.free\_required} field indicates whether the data stored in the \code{cdt.cdt\_val} field needs to be freed.

The {cdt.cdt\_val} field  is of type\code{cdt\_types} which is a union presented in the listing \ref{code:cdt_types}. \code{cdt\_types} valid field is based on the value of \code{cdt.type}.

\begin{listing}[h]
\begin{singlespace}
\begin{minted}[bgcolor=CodeBG,fontsize=\footnotesize,autogobble,tabsize=2,linenos]{c}
union cdt_types
{
	metaffi_float32 float32_val;
	metaffi_float64 float64_val;
	metaffi_int8 int8_val;
	metaffi_uint8 uint8_val;
	metaffi_int16 int16_val;
	metaffi_uint16 uint16_val;
	metaffi_int32 int32_val;
	metaffi_uint32 uint32_val;
	metaffi_int64 int64_val;
	metaffi_uint64 uint64_val;
	metaffi_bool bool_val;
	struct metaffi_char8 char8_val;
	metaffi_string8 string8_val;
	struct metaffi_char16 char16_val;
	metaffi_string16 string16_val;
	struct metaffi_char32 char32_val;
	metaffi_string32 string32_val;
	struct cdt_metaffi_handle* handle_val;
	struct cdt_metaffi_callable* callable_val;
	struct cdts* array_val;
 };
\end{minted}
\end{singlespace}
\caption{cdt\_types union}
\label{code:cdt_types}
\end{listing}

\subsection{MetaFFI Types}\label{sec:metaffi-type}

CDT supports 24 data types detailed in table \ref{tab:metaffi_types}.
\begin{table*}[h]
    \begin{tabular}{| c c c c |}
        \hline
        \rowcolor{lightgray}\multicolumn{4}{|c|}{Numeric Types} \\
        \hline
        \hline
        float32 & float64 & & \emph{bool} \\
        int8 & int16 & int32 & int64 \\
        uint8 & uint16 & uint32 & uint64 \\
        \hline
    \end{tabular}
    \begin{tabular}{| c c |}
        \hline
        \rowcolor{lightgray} \multicolumn{2}{|c|}{String Types} \\
        \hline
        \hline
        char8 & string8 \\
        char16 & string16 \\
        char32 & string32 \\
        \hline
    \end{tabular}
    \begin{tabular}{| c c |}
        \hline
        \rowcolor{lightgray} \multicolumn{2}{|c|}{Special Types} \\
        \hline
        \hline
        handle & callable\\
        any &\\
        & \\
        \hline
    \end{tabular}
    \begin{tabular}{| c c |}
        \hline
        \rowcolor{lightgray} \multicolumn{2}{|c|}{Internal Types} \\
        \hline
        \hline
        null &  array\\
        size &  type\\
        &\\
        \hline
    \end{tabular}
    \caption{MetaFFI Types}
    \label{tab:metaffi_types}
\end{table*}

\subsubsection{Numeric Types}

The \emph{numeric} types are passed by value. When there is no equivalent numeric data type in the target runtime, the plug-in implementer has to choose a suitable conversion. For instance, Java does not support unsigned 64-bit integers, so the plug-in can map them to signed int64 or use Java BigInteger. Alternatively, the plug-in implementer can indicate that a certain type is not supported by returning an error, but this is not advisable.

The structure \code{cdt\_T} represents a numeric type \textit{T} and is defined as follows:
\begin{minted}[bgcolor=CodeBG,fontsize=\footnotesize,autogobble,tabsize=2,linenos]{c}
struct cdt_T
{
    T val; // The value of type T
};
\end{minted}


\subsubsection{String Types}

Like \emph{numeric types}, \emph{string types} are used to pass parameters and return values to the equivalent data type at the target runtime. While numeric types are always copied, string types might get copied, depending on the caller runtime support of C-String. In case the caller runtime can pass a null-terminated C pointer of the right encoding to the string parameter, there is no need to copy the string. Notice that if the parameter is considered an in-out or out-parameter, the caller plugin must take it into consideration if the string should be copied or not.

The structure \code{cdt\_STRING\_T} represents a string type \textit{STRING\_T*} and is defined as follows:
\begin{minted}[bgcolor=CodeBG,fontsize=\footnotesize,autogobble,tabsize=2,linenos]{c}
struct cdt_STRING_T
{
	STRING_T* val; // null terminated string
};
\end{minted}



\subsubsection{Handle Type}

In MetaFFI, a handle is characterized as follows: If the handle is associated with the current runtime, it returns the original object; otherwise, it returns the structure \code{cdt\_metaffi\_handle*}.

The structure \code{cdt\_metaffi\_handle} represents a handle:
\begin{minted}[bgcolor=CodeBG,fontsize=\footnotesize,autogobble,tabsize=2,linenos]{c}
typedef void* metaffi_handle;
struct cdt_metaffi_handle
{
	metaffi_handle handle; // The void* representing the object
	uint64_t runtime_id; // The runtime the handle originated from
    void* release; // void((*)(cdt_metaffi_handle*)) C-function which releases the entity
};
\end{minted}

A \emph{handle} is a \code{void*} that refers to an entity from another language at runtime. The \code{void*} is not necessarily a valid pointer, but a representation of the object which its runtime can use to retrieve the original object. To determine the runtime associated with a \emph{handle}, each runtime is assigned a distinct \emph{runtime\_id}, whose 64-bit integer is generated randomly and embedded in the runtime plugin. When a runtime registers an entity using the \code{cdt\_metaffi\_handle} struct, it assigns \emph{handle} and \emph{runtime\_id} to the struct.

The release field points to a function that takes \code{cdt\_metaffi\_handle} and releases the object. It is important to note that a plugin developer should release the object after a function returns. Instead, the release function, which manages the object's lifetime, is defined by the user. Releasing the object does not mean that the object is deleted, as the object's original runtime determines when the object is actually being deleted.



The plugin implementer is responsible to map the \code{void*} to the entity at runtime. As an important note for implementers, in managed memory runtimes (e.g. Java, Python, Go), it is important to make sure that the managed entity is not deleted when used by an external runtime. MetaFFI runtime plugin must prevent the entity from being deleted until the \code{release} function in the \code{cdt\_metaffi\_handle} is called.

Two simple examples of implementing a handle is in Python3.11 and JVM. In Python3.11, all objects are represented by pointers to a \code{PyObject} structure, so the handle value is simply a \code{PyObject*}. Python employs a reference counting scheme \cite{cpython_garbage_collection} to manage memory, so to prevent the object from being garbage collected, we need to increase the reference count using \code{Py\_INCREF} when creating the handle, and decrease it using \code{Py\_DECREF} when releasing the handle. In JVM languages, all objects are represented by pointers to a \code{\_jobject} structure, so the handle value is simply a \code{\_jobject*}. To prevent the object from being garbage collected \cite{jvm_garbage_collector}, we need to create a global reference using \code{NewGlobalRef} when creating the handle and delete it using \code{DeleteGlobalRef} when releasing the handle.

In Go, the situation is more complicated, as there is no C-compatible data type that can represent any Go object. In this case, we use a global object table, where the key is a random 64-bit unsigned integer, and the value is a Go object stored as an empty interface \code{interface{}}, which can hold any Go type. The handle value is then the key to the object table entry. Go uses mark and sweep garbage collection \cite{go_memory_management}, but there is no direct method to mark an object. Therefore, as long as the object is in the table, the Go runtime will not collect it. The release function removes the entry from the table, thus freeing the reference to the object. Note that this technique requires using synchronization primitives for mutual exclusion, such as a readers/writer lock.

\subsubsection{Any Type}
The \emph{any} type indicates that the CDT type is dynamic and determined in runtime.  The CDT field \code{type} specifies the actual type.

\subsubsection{Array Type} \label{sec:types_array}
In programming languages we can find several types of arrays:
\begin{itemize}
    \item  Classic array - fixed dimensions with the same length\\
    example: \code{[1,2][3,4]}
    \item  Ragged array - fixed dimension with different length\\
    example: \code{[1][2,3][4,5,6]}
    \item  Mixed dimensions array - every element might be an array of different dimension\\
    example: \code{1,[2,3][[4,5][6,7]]}
\end{itemize}
As MetaFFI tries to support all scenarios, arrays in CDT support mixed dimensions array (similar to lists in Python) which supports all types of arrays. Every array in CDT has an array header of the CDTS structure, which can be held by CDT in the \code{cdt\_types} union. The presented CDTS structure holds an array of CDT. Therefore, every element can be either a data type or an array.

\begin{minted}[bgcolor=CodeBG,fontsize=\footnotesize,autogobble,tabsize=2,linenos]{c}
struct cdts
{
	struct cdt*     arr;
	metaffi_size    length;
 }
 \end{minted}

\subsubsection{Callable Type}

The \emph{callable} type represents a C function that invokes MetaFFI XCall. The structure \code{cdt\_metaffi\_callable} is used to pass an XCall as an argument or return it as a result, as illustrated in Code Listing \ref{code:py3-callable}, where the Java method \code{add} is passed to the Python function \code{call\_callback\_binary\_op} as a parameter and is called back from within Python.

The structure \code{cdt\_metaffi\_callable} contains the information required for a runtime plugin to create the proper CDT for a pointer to the C function that performs the XCall.

\begin{minted}
[bgcolor=CodeBG,fontsize=\footnotesize,autogobble,tabsize=2,linenos]{c}
struct cdt_metaffi_callable
{
	struct xcall* val;

    // the definition of the XCall
	metaffi_type* parameters_types; // array of MetaFFI types
	metaffi_int8 params_types_length; // length of parameters_types
	metaffi_type* retval_types; // array of MetaFFI types
	metaffi_int8 retval_types_length; // length of retval_types
};
\end{minted}

The field \code{val} stores a pointer to the XCall structure which holds the C function pointer to the XCall and context. We further discuss xcall structure in section \ref{sec:xcall}. The \code{parameter\_types}, \code{retval\_types}, and their respective lengths specify the signature of the XCall, similar to the types declared when loading a foreign entity using the \code{load\_entity} function.

\subsubsection{Null Type}
The \emph{null} type indicates \emph{null} or \code{void} type defining the foreign entity signature for the \code{load\_entity} or \code{make\_callable} functions. It is similar to passing a \code{NULL} pointer to indicate that there are no parameters or return values.

\subsubsection{Size and Type Types}
The \emph{size} and \emph{type} types are not CDT types in MetaFFI, but 64-bit unsigned integers that are used as \code{typedef} throughout the MetaFFI system. The \emph{size} type is a data type for storing sizes and the \emph{type} type is a data type for storing MetaFFI types.

\subsection{XCall - Cross Language} \label{sec:xcall}

The functions \code{load\_entity} and \code{make\_callable} return an XCall structure, which is constructed of two elements: C function pointer to the XCall entry point, and a pointer to a runtime plugin specific context structure.

\begin{minted}
[bgcolor=CodeBG,fontsize=\footnotesize,autogobble,tabsize=2,linenos]{c}
struct xcall
{
	void* pxcall_and_context[2];
};
\end{minted}

The XCall entry point is a C function pointer to:
\begin{minted}[bgcolor=CodeBG,fontsize=\footnotesize,autogobble,tabsize=2,linenos]{c}
void (*)(void* context, cdts* pcdts, char** out_err);
\end{minted}
In case the foreign entity does not require parameters and has no return values the C function pointer is:
\begin{minted}[bgcolor=CodeBG,fontsize=\footnotesize,autogobble,tabsize=2,linenos]{c}
void (*)(void* context, char** out_err,);
\end{minted}

The parameter \code{context} is the second element returned by the functions \code{load\_entity} and \code{make\_callable}. It stores the information that the runtime plugin needs to invoke and use the foreign entity.  The functions are implemented in the runtime plugin, and create the context. If the runtime plugin does not need any context, the context pointer is \code{null}.

In the entrypoint function, the parameter \code{pcdts} refers to two CDTS structures, where \code{pcdts[0]} contains the CDTS for the arguments and \code{pcdts[1]} contains the CDTS for the return values. If there are no arguments or results, the \code{len} field in the CDTS structure is set to 0.

The parameter \code{out\_err} is an \emph{out} parameter that stores a null-terminated error string, in the event of an error. The caller is responsible for initializing the parameter to \code{null}. As discussed in section \ref{sec:mem_alloc_dealloc} the error string must be allocated using XLLR memory allocation function. In case of an error, the caller must free the error using the XLLR free function. Therefore, the runtime plugin must allocate the error string on the heap. This can be improved in the future by adding another \emph{out} Boolean parameter \code{is\_free\_err}.

\subsubsection{CDTS allocation}
CDT is being reused in every MetaFFI XCall, therefore its allocation and deallocation affect the performance of the system.

MetaFFI does not assume that the calling runtime can allocate CDT on the stack. Also, MetaFFI cannot also provide a C function that allocates CDT on the stack because the CDT would be freed once the C function returns.

To overcome this, MetaFFI provides the CDT allocation and deallocation function:
\begin{minted}
[bgcolor=CodeBG,fontsize=\footnotesize,autogobble,tabsize=2,linenos]{c}
cdts** alloc_cdts_buffer(metaffi_size params_count, metaffi_size ret_count);
void free_cdts_buffer(cdts** pcdts);
\end{minted}

Instead of allocating CDTS and CDTs on the heap every call, MetaFFI pre-allocates 50 CDTS and 50 CDTs globally, per thread, using thread local storage \cite{ThreadLocalStorage} and index for each type to specify the current offset of used CDTS and CDTs.
\begin{minted}
[bgcolor=CodeBG,fontsize=\footnotesize,autogobble,tabsize=2,linenos]{c}
thread_local cdt cdt_cache[cdt_cache_size];
thread_local int cdt_current_index = 0;
thread_local cdts cdts_cache[cdts_cache_size];
thread_local int cdts_current_index = 0;
\end{minted}

The function \code{alloc\_cdts\_buffer} reuses CDTS and CDTs from a global cache if the number of CDTs required for the call is within a predefined threshold, otherwise it dynamically allocates them on the heap and updates the flag \code{allocated\_on\_stack} in the CDTS structure accordingly. The threshold is set at 50, which means that the function can handle up to 50 CDTS and CDTs without allocating memory (i.e. \code{params\_count}+\code{ret\_count} $\leq$ 50). The function also updates \code{cdt\_current\_index} based on the number of parameters and return values, also adding two to \code{cdts\_current\_index}, since every call requires two CDTS.

Based on the value of \code{allocated\_on\_stack}, the function \code{free\_cdts\_buffer} frees the CDTS and CDTs from the heap or decrements the fields \code{cdt\_current\_index} and \code{cdts\_current\_index}.

\subsubsection{Cleanup}
The \code{cdt\_val} field of a CDT may contain data that is dynamically allocated to the heap by a runtime plugin. In such cases, the runtime plugin is required to set the \verb|free_required| field of the CDT to TRUE (a nonzero value). This indicates that the caller needs to free the data after the call using the XLLR free function.

\subsection{Capability Based XCall Optimization\protect\footnote{This section is not yet implemented in MetaFFI}} \label{sec:cdt_req}

Various programming languages provide different capabilities that can influence the efficiency of XCall (i.e., interoperability). Typically, we only require C-FFI for programming languages, which informs the present design of the CDT. However, if certain assumptions about the matching programming languages can be made, it can enhance the performance of XCall by using different entries in the \code{cdt\_types} union within CDT that would not be otherwise used.

In order to match a pair of programming languages and decide which fields in the CDT union they are using, the function \code{load\_runtime} returns an unsigned 64-bit integer specifying the available capabilities of the runtime plugin. The loading runtime checks these capabilities, and if they match, the calling may choose using different fields in \code{cdt\_types} which are more efficient. When calling the functions \code{load\_entity} or \code{make\_callable}, the caller also passes on the capabilities it wants to use, based on the capabilities returned in \code{load\_runtime}.

For example, an array of integers is passed within an array of CDT structures, each of them stores an integer. In the general case, the generic design is required as each element can hold a different type, or even another array. But, if we are calling from a statically typed language like Go to a statically typed language like Java or C++, we can pass a pointer to a C array of integers using a dedicated field in \code{cdt\_types} and setting the correct type, as we know that both runtimes expect integers. 

The mechanism can even skip CDT altogether, if both runtimes support it. For example, assume that we are calling from C++ to C, and both of them support the x86-64 calling convention. Once C++ calls \code{load\_runtime}, the x86-64 calling convention flag is set in the returned capabilities. When the C++ client loads a foreign entity using \code{load\_entity}, it passes the x86-64 calling convention capability. In turn, the C runtime plugin returns a pointer to a C function expecting x86-64 calling convention which forwards the call to the foreign entity.

Using capabilities, XCall and CDT can enhance the performance based on assumptions that can be determined during runtime. 

It is important to note that the capabilities are meant for optimization. In order for a runtime plugin to support all programming languages supporting MetaFFI, the plugin must implement the general case of CDT and XCall. Once the general case is implemented, the plugin can use capabilities to optimize its performance with specific types of runtime.

\subsection{Compiler \& Compiler Plugin}\label{sec:compiler}

\subsubsection{Guest Compiler} \label{sec:guest-compiler}

Many runtimes support the functionality of loading entities from their respective "executable-code" formats. For example, JVM can load entities from \code{.jar} or \code{.class} files, while CPython can load entities from \code{.py} or \code{.pyc} files. However, this functionality is not uniformly supported in all runtimes, and some runtimes, such as \code{Go}, require the creation of entry points to access foreign entities.

To address this challenge, MetaFFI provides a compiler that generates "executable code" for the target runtime with entry points to the foreign entities.

The \code{metaffi} command line tool provides a command line to execute the compiler:
\verb|metaffi -c --idl [path] -g|

The options are as follows:
\begin{itemize}
\item \textbf{-c}: compile
\item \textbf{--idl}: path to source code to extract signatures from foreign entities
\item \textbf{-g}: generate ``guest code''
\end{itemize}

For instance, to build a guest module for \code{TestRuntime.go} execute in the terminal:
\verb|metaffi -c --idl TestRuntime.go -g|

The command produces a dynamic library that the user can load as a MetaFFI module and access the foreign entities.

A \emph{guest compiler} plugin is a dynamic library that implements the following C interface\footnote{In future versions of MetaFFI, a plugin will be loaded using MetaFFI, therefore it could be implemented in any supported language. In case the loading fails, MetaFFI will revert to the C interface}:

\begin{minted}
[bgcolor=CodeBG,fontsize=\footnotesize,autogobble,tabsize=2,linenos]{c}

// called once when the compiler plugin is loaded to perform any initialization tasks.
void init();

// generates executable for the target runtime based on given JSON IDL
void compile_to_guest(const char* idl_def_json,
						   const char* output_path, 
						   const char* guest_options, 
						   char** out_err);
 \end{minted}

The \code{compile\_to\_guest} function is invoked by the MetaFFI command-line tool to generate a MetaFFI module for the target runtime with entry points for its entity. The function takes as input an IDL definition, which is a JavaScript Object Notation (JSON) formatted string \cite{json_rfc} that specifies the signatures of all foreign entities in the target code to be available via MetaFFI. The IDL and its automatic generation are explained in section \ref{sec:metaffi_idl_plugin}.

The function has the following parameters (strings are null terminated):
\begin{itemize}
\item \verb|const char* idl_def_json|: the IDL to compile
\item \verb|const char* output_path|: the path where the module should be written
\item \verb|const char* guest_options|: an optional key-value string in the format key1=value1,...,keyN=valueN that provides options for the plugin compiler. Options are passed using the MetaFFI command line switch \verb|--guest-options|
\item \verb|char** out_err|: an output parameter that returns an error string in case of failure
\end{itemize}

For compiler plugin implementers, the error string returned by \verb|out_err| must be allocated using the XLLR memory allocation function, as the MetaFFI tool will attempt to free it using the XLLR free function. Also, in case the runtime supports access control for entities (e.g. \emph{public}, \emph{private}), it is recommended to generate entry points only for public entities. 

\subsubsection{Host Compiler}\label{sec:host-compiler}

The goal of the host compiler plugin is to automatically generate code that takes advantage of the language-specific MetaFFI API to load and wrap foreign entities. The generated code enables transparent use of the foreign entities, satisfying the \emph{simple interoperability}\cite{empirical_multi_lingual} requirement (presented in section \ref{sec:intro}), which states that the code accessing foreign entities should be the same as accessing entities within the runtime.

The host compiler can be essential when using a large foreign library, as it frees the developer from writing code to load the foreign entities, which can be an exhausting task for large libraries.

A host compiler plugin, analogous to a guest compiler plugin, is a dynamic library that implements the following C functions\footnote{Similar to a guest compiler, in the future a host compiler will first attempt to load as a MetaFFI module.}:

\begin{minted} [bgcolor=CodeBG,fontsize=\footnotesize,autogobble,tabsize=2,linenos]{c}

// called once when the compiler plugin is loaded to perform any initialization tasks.
void init();

// generates code for the required runtime which uses MetaFFI API to load
// and call the foreign entities
void compile_from_host(const char* idl_def_json,
                                            const char* output_path,
                                            const char* host_options,
                                            char** out_err); 
\end{minted}

The \code{compile\_from\_host} function is invoked by the MetaFFI command-line tool to generate source code that uses the MetaFFI API to load and use foreign entities. The function takes as input an IDL definition that specifies the signatures of all the entities in the target code. The IDL and its automatic generation are explained in section \ref{sec:metaffi_idl_plugin}.

The function has the following parameters:
\begin{itemize}
    \item \verb|const char* idl_def_json|: the IDL to compile
    \item \verb|const char* output_path|: the path where the module should be written
    \item \verb|const char* host_options|: similar to guest-options, an optional key-value string in the format key1=value1,…,keyN=valueN that provides options for the plugin compiler. Options are passed using the MetaFFI command line switch \verb|–host-options|
    \item \verb|char** out_err|: an output parameter that returns an error string in case of failure 
\end{itemize}

Similarly to the guest-compiler plugin, the error string returned by \verb|out_err| must be allocated using the XLLR allocation function.

\subsection{IDL Definition Model \& IDL Plugin} \label{sec:metaffi_idl_plugin}\label{sec:def-model}

The IDL, which defines the signature and structure of foreign entities, is a hierarchical structured JSON format string required by MetaFFI Compiler plugins (section \ref{sec:compiler}). The IDL provides the information necessary for the guest compiler plugins to generate code that interacts with the foreign entities, and the host compiler to wrap the foreign entities. The IDL specifies the different entities and their properties. Section \ref{sec:idl-details} contains detailed and technical information on the IDL.

\begin{figure}[t]
  \centering
  \includegraphics[scale=0.3]{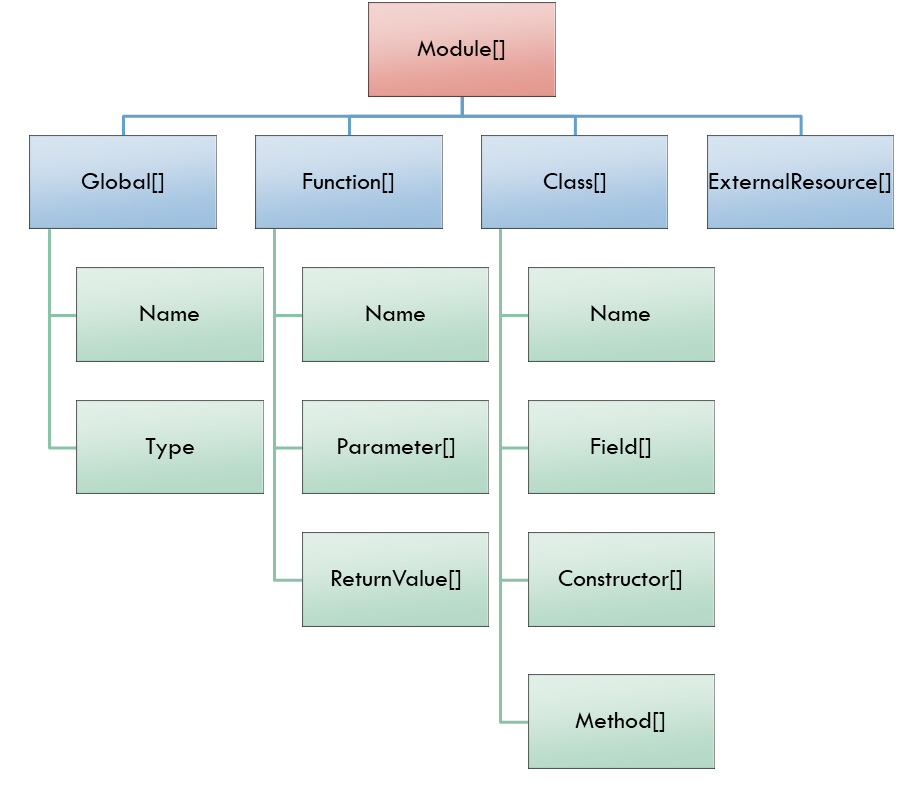}
  \caption{Simplified Definition Model}
  \label{fig:simplified-def-model}
\end{figure}

As shown in figure \ref{fig:simplified-def-model}, which illustrates a simplified view of the definition model, the IDL consists of modules that correspond to the foreign modules. Each module contains four different entities: 
\begin{itemize}
    \item Globals, which are global variables and constants
    \item Functions, which are global functions or global scripts \footnote{Global script can be represented as a function named \code{main}}
    \item External Resources, which are external modules and resources required by the guest compiler
    \item Classes, which are complex types, such as \code{class} (as found in C++, Java, Python3.11 etc.), \code{struct} (as exist in C, Go etc.) and \code{record} (in Emacs Lisp 26.1 \cite{emacs-lisp-26-1})
    \begin{itemize}
    \item Fields, which are the attributes of the complex types
    \item Constructors are used to create new instances of complex types, similar to constructors in C++ or Java. In cases where the runtime does not support constructors, the implementer may choose to represent certain functions as constructors, as long as they still return a new instance of the complex type
    \item  Methods, which can also be functions in runtimes that do not support methods in case they accept an instance of the complex type as a parameter. The developer has the flexibility to define certain functions as methods, as long as they necessitate an instance of the complex type
    \end{itemize}
\end{itemize}

MetaFFI provides an implementation of the IDL entities and creates a JSON IDL.

\subsubsection{IDL Plugin}

Kaplan et al.  \cite{polispin} stated, writing IDL manually is a tedious and impractical approach with many drawbacks. As we agree with this statement, MetaFFI generates the IDL automatically and does not require any manual writing of the IDL. Instead, MetaFFI uses IDL plugins to extract IDL directly from source code or runtime execution code. Automated IDL generation also satisfies \emph{simple interoperability} \cite{empirical_multi_lingual} as presented in section \ref{sec:intro} .

The IDL plugin analyses source code or runtime executable code and extracts the definitions of the foreign entities (excluding the actual logic of the foreign callable entities) and constructs the definition data using the entities of the definition model.

Like other MetaFFI plugins, the IDL plugin is a dynamic library that exports the following C functions\footnote{Similar to the compiler plugins, the IDL plugin will attempt to load as a MetaFFI module in the future.}. IDL compiler need to implement the following function:

\begin{minted} [bgcolor=CodeBG,fontsize=\footnotesize,autogobble,tabsize=2,linenos]{c}
// called once when the idl plugin is loaded to perform any initialization tasks.
void init();

enum idl_input_type
{
    source_code = 0,
    path = 1
};

// generates MetaFFI IDL from a given source code, file or path
char* parse_idl(idl_input_type input_type,
						 const char* data,
						 char** out_err);
\end{minted}

\begin{itemize}
    \item \code{input\_type} the type of data passed in \code{data}
    \item \code{data} contains a file or path of the source code or executable-code to parse.
    \item \code{out\_err} contains an error string, in case of an error. The error string must be by XLLR allocation function.
    \item \code{return} the generated IDL in JSON format. The IDL must be stored on the heap using XLLR allocation function.
\end{itemize}

\subsection{MetaFFI Supported Languages} \label{sec:lang_prereq}

MetaFFI focuses on \emph{programming languages}, which are languages used to write the logic of the application, as defined in \cite{empirical_multi_lingual}. This differs from non-programming languages that are designed to store data, describe a web page, and so on (e.g. HTML \cite{html_rfc}, GNU Make \cite{makefile}, CSS \cite{css_rfc}, JSON \cite{json_rfc}, XML \cite{xml_rfc}).

MetaFFI does not limit itself to \emph{programming languages}, but also supports a broad definition of “entities”, which can be found outside the programming languages domain. For instance, SQL stored procedures \cite{sql_stored_procedure}. Even though SQL is typically not a programming language used to write logic of application, and even executed outside the process that “calls” the SQL code, MetaFFI can be used to invoke stored procedures as if they were functions in the host language.

Table \ref{tbl:supported-languages} lists a few languages and the prerequisites they satisfy.
\begin{table*}[t]
  \centering
\begin{tabular}{ | c || c | c | }
    \hline
    \rowcolor{lightgray} \emph{Languages} & \emph{Act as Host} & \emph{Act as Guest} \\
    \hline
    \hline
    Go\protect\footnotemark[6] & \checkmark  & \checkmark \\
    \hline
    {Python3.11}\protect\footnotemark[6] & \checkmark & \checkmark \\
    \hline
    OpenJDK\protect\footnotemark[6] & \checkmark & \checkmark  \\
    \hline
    .Net Framework & \checkmark & \checkmark  \\
    \hline
    C & \checkmark & \checkmark  \\
    \hline
    C++ & \checkmark & \checkmark \\
    \hline
    LLVM-IR\protect & \checkmark & \checkmark  \\
    \hline
    JavaScript NodeJS & \checkmark & \checkmark  \\
    \hline
    Bash & \checkmark & \checkmark  \\
    \hline
    Powershell & \checkmark & \checkmark  \\
    \hline
    JavaScript V8 Engine & \checkmark & \checkmark \\
    \hline
    SQL & & \checkmark  \\
    \hline
    In Browser JavaScript & & \\
    \hline
    HTML & & \\
    \hline
\end{tabular}
  \caption{Partial list of MetaFFI language supported}
  \label{tbl:supported-languages}
\end{table*}

\section[Implementer Guide - How To Add a Language Support?]{Implementer Guide - How To Add a Language Support? \technical} \label{sec:add_lang}

This section describes the process of adding new language support and provides step-by-step instructions on how to implement the various plugins. Let $L$ be the new language that we want to add.

Before diving into the details, notice that to support calls to foreign entities, make sure $L$ can call \code{C} functions from a dynamic library, and call C-function pointer.\\
To support other languages calling $L$, make sure that $L$ can be called from \code{C}.\\

\subsection{$L$ as a Host Language}

In order for $L$ to interact with other languages via MetaFFI, it should interact with the XLLR API, as described in section \ref{sec:xllr}\footnote{MetaFFI plugin SDK found in MetaFFI GitHub repository provides more helper functions}:

\begin{minted}[bgcolor=CodeBG,fontsize=\footnotesize,autogobble,tabsize=2,linenos]{c}
void load_runtime_plugin(const char* runtime_plugin_name, 
                                                 char** out_err);
                                            
void free_runtime_plugin(const char* runtime_plugin,
                                            char** out_err);
                                            
void** load_function(const char* runtime_plugin_name,
                                        const char* module_path,
                                        const char* function_path,
                                        metaffi_type_info* params_types, int8_t params_count,
                                        metaffi_type_info* retval_types, int8_t retval_count,
                                        char** out_err);

// allocates CDTS
pcdts** alloc_cdts_buffer(uint64_t params_count, uint64_t ret_count);

// free CDTS (allocated with alloc_cdts_buffer)
void free_cdts_buffer(pcdt*);

// allocate memory on the heap
void* metaffi_alloc(uint64_t size);

// free memory (allocated with metaffi_alloc)
void metaffi_free(void* ptr);

// string allocation functions and free function
// mainly assist to allocate strings returned as errors
char* alloc_string(const char* err, uint64_t length);
char8_t* alloc_string8(const char8_t* err, uint64_t length);
char16_t* alloc_string16(const char16_t* err, uint64_t length);
char32_t* alloc_string32(const char32_t* err, uint64_t length);
void free_string(const char* err_to_free);
\end{minted}

\begin{figure}[h]
    \centering
    \includegraphics[width=\linewidth]{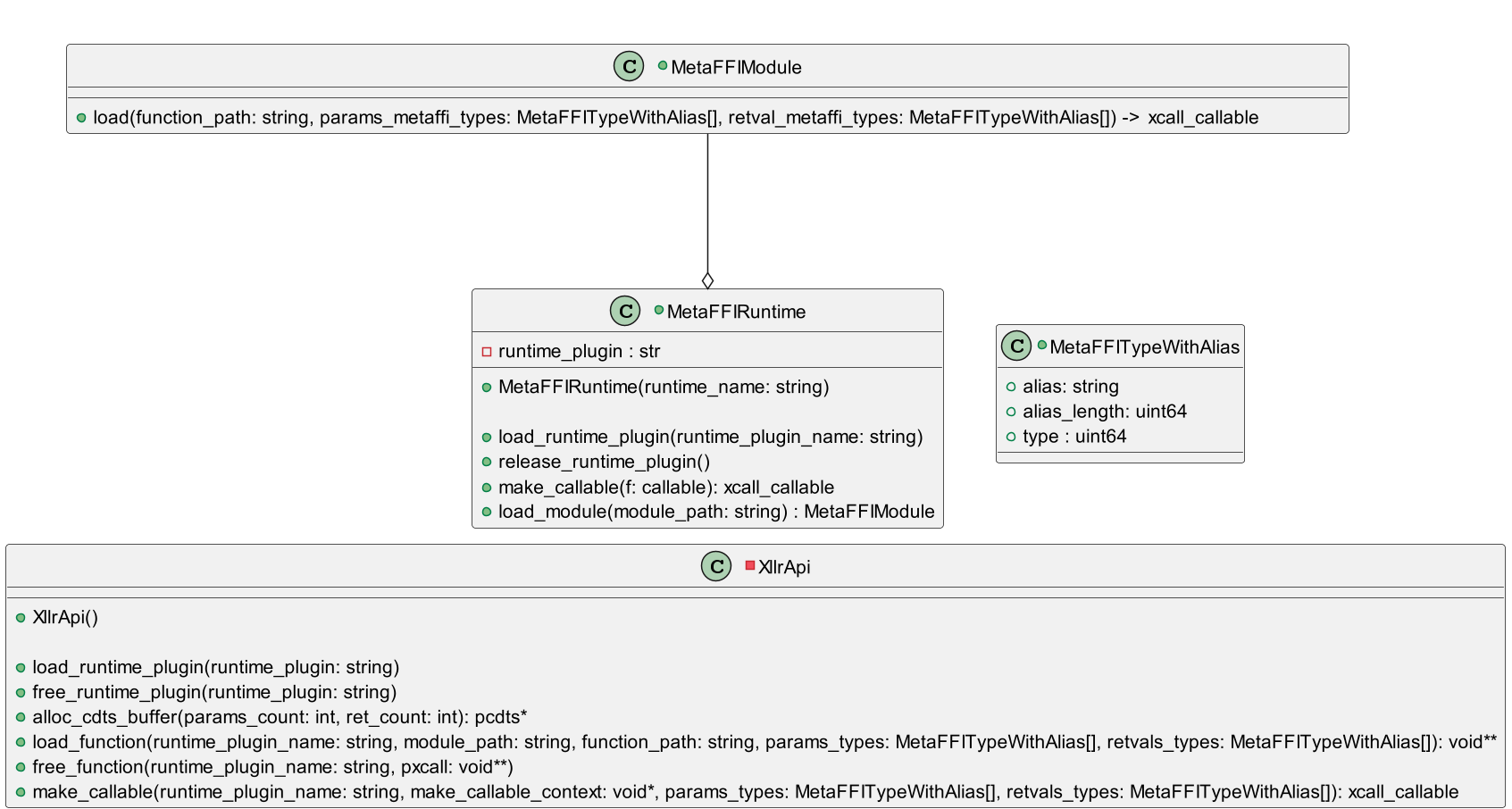}
    \caption{Class diagram for generic MetaFFI API of $L$}
    \label{fig:l-api}
\end{figure}

We recommend wrapping the XLLR API by implementing an API in the $L$ source code. The general design of the API is shown in figure \ref{fig:l-api}.

\code{MetaFFIRuntime} class represents the MetaFFI runtime plugin, which is responsible for loading and releasing the runtime and managing modules of that runtime. The \code{MetaFFIRuntime} class has a constructor that takes a string argument representing the name of the runtime plugin. The \code{MetaFFIRuntime} class also has the following functions:

\begin{itemize}
    \item \verb|load_runtime_plugin|: This method loads the runtime associated with the runtime plugin.
    \item \verb|release_runtime_plugin|: This method releases the runtime used by the plugin.
    \item \verb|load_module|: This method creates and returns a \code{MetaFFIModule} object.
\end{itemize}

The \code{MetaFFIModule} class, using the \code{load} method, is responsible for loading foreign entities from a given function path and an array of \code{MetaFFITypeInfo} objects for the parameters and the return value of the function. The method returns a callable object that wraps the XCall. The \code{MetaFFITypeInfo} class encapsulates the C-struct \code{metaffi\_type\_info}.

The \code{make\_callable} method in the \code{MetaFFIRuntime} class takes a callable object $f$ in language L and uses the \code{make\_callable} function of XLLR to return an XCall for $f$ that can be passed to other foreign entities. Note that \code{make\_callable} can only be supported if $L$ is also a guest language. Before calling to \code{make\_callable}, the runtime plugin of $L$ must be loaded.

The returned callable object (or an object with a \code{call} method) invokes the XCall C-function pointer stored as the first element in the returned \code{XCall} structure from \code{load\_entity}. The XCall entry point accepts and returns values according to the MetaFFI Types defined in the \code{load\_entity}. The callable object accepts and returns values as defined by the \code{load} method. This callable object enables the language $L$, using the API, to handle calling foreign entities with different signatures and types in a uniform way.

In case of an error returned from the XLLR functions or XCall, the system should return or throw an error using the standard methodology common in language $L$, such as exceptions or error values, and free the error string using the \code{metaffi\_free} function.

\subsubsection{Implementing Host Compiler} \label{sec:l-host-compiler}

The automatic generation of $L$ source code using the API requires the development of a Host Compiler. It is crucial to note that when employing $L$ as the host compiler to generate code for the language $L'$, an IDL plugin must be implemented for $L'$. The implementation of the IDL plugin is detailed in section \ref{sec:l-idl-plugin}.

To implement the host compiler, a dynamic library must be created that exports the following functions\footnote{In the future, host compiler as a MetaFFI module will be supported}:
\begin{minted} [bgcolor=CodeBG,fontsize=\footnotesize,autogobble,tabsize=2,linenos]{c}
// called once when the compiler plugin is loaded to perform any initialization tasks.
void init();

// called to compile the guest code stated by the IDL definition to the host source code.
void compile_from_host(const char* idl_def_json, uint32_t idl_def_json_length,
                                const char* output_path, uint32_t output_path_length,
                                const char* host_options, uint32_t host_options_length,
                                char** out_err, uint32_t* out_err_len); 
\end{minted}

The function \code{compile\_from\_host} receives the IDL as a JSON string and an output path that specifies where to write the source code. The MetaFFI tool can also receive host compiler options in the format of $key_i=value_i,key_j=value_j$. These host options are passed to \code{compile\_from\_host} in the \code{host\_options} parameter. The function should generate $L$ source code, using the $L$ MetaFFI API to load $L'$ runtime and provide $L$ wrappers to $L'$ foreign entities.

The MetaFFI Software Development Kit (SDK) offers \code{Go} code for producing compilers. Go was selected due to its robust template package, which is beneficial for generating code.

The following code represents a plugin that utilizes the SDK to compile a dynamic library for the host:
\begin{minted}[bgcolor=CodeBG,fontsize=\footnotesize,autogobble,tabsize=2,linenos]{go}
// PluginMain.go
package main

import "C"

import "github.com/MetaFFI/plugin-sdk/compiler/go"

//export init_plugin
func init_plugin() {
	compiler.PluginMain = compiler.NewCompilerPluginMain(NewHostCompiler(), NewGuestCompiler())
}
func main() {}
\end{minted}

This code snippet from \code{PluginMain.go} uses the \code{NewCompilerPluginMain} function imported from the plugin SDK. It ensures that the created Go dynamic library exports the correct functions.

The \code{NewCompilerPluginMain} function expects an instance of a host compiler and an instance of a guest compiler. If you do not wish to implement a guest compiler, implement the \code{Compile} method for the Guest Compiler (as detailed in section \ref{sec:l-guest-compiler}), but simply return a "Not Implemented" error.

The host compiler is a Go struct implementing the \code{Compile} method:
\begin{minted}[bgcolor=CodeBG,fontsize=\footnotesize,autogobble,tabsize=2,linenos]{go}
type HostCompiler struct {}
func (c *HostCompiler) Compile(definition *IDL.IDLDefinition, 
                                outputDir string,
                                outputFilename string,
                                hostOptions map[string]string) (err error)
\end{minted}

The method provides the IDL JSON string as IDL Go entities (detailed in \ref{sec:idl-details}). In addition, the host options, if any, have been parsed into a \code{map}.

In MetaFFI, the Go \code{text/template} package is used to implement code generation by creating repeated patterns to the $L$ source code based on the entities in the given IDL.

You can find the Go host compiler and host template in the official MetaFFI GitHub repository.

\subsection{$L$ as a Guest Language} \label{sec:l-guest-language}

Prior to implementing the guest support, it is imperative to investigate several aspects related to $L$:
\begin{itemize} 
    \item In the event that it is not feasible to load $L$ runtime executable modules and access the entities (e.g. Go), the following steps should be undertaken:
        \begin{itemize}
            \item Implement an IDL plugin
            \item Implement a guest compiler plugin
        \end{itemize}
    \item Implement a runtime plugin
    \begin{itemize}
        \item Mapping of MetaFFI numeric types to the corresponding data types in $L$. If there is a reluctance to support data types, although it is not advisable, a runtime error may be raised with "unsupported type".
        \item Mapping of MetaFFI string types to data types in $L$. If there is a disinclination to support all types of string, although it is not advisable, a runtime error may be raised with "unsupported type".
        \item For the support of complex types - it is necessary to research how to prevent entities from being deleted, and how to delete them post usage.
        \item To facilitate the passing of $L$ callable as a parameter or return value - it is crucial to research how to represent or pass callable of $L$ to C.
    \end{itemize}
\end{itemize}

XLLR invokes the runtime plugin to manage $L$'s runtime, manage entities in $L$ and encapsulate $L$ entities with XCall, thereby making it accessible from other languages. Moreover, in many instances, the runtime plugin contains the entry point for the XCall (unless a guest compiler is necessitated). Implementing XCall is detailed in section \ref{sec:l-xcall}.

The runtime plugin is a dynamic library that exports the following C functions:
\begin{minted}[bgcolor=CodeBG,fontsize=\footnotesize,autogobble,tabsize=2,linenos]{c}
// Load runtime 
void load_runtime(char** err);

// Free runtime
void free_runtime(char** err);
 
// Load entity
XCall* load_entity(const char* module_path,
            const char* function_path,
            metaffi_type_info* params_types, 
            int8_t params_count,
            metaffi_type_info* retvals_types,
            int8_t retval_count,
            char** err);

// Wrap callable with XCall
void** make_callable(void* make_callable_context,
            metaffi_type_info* params_types,
            int8_t params_count,
            metaffi_type_info* retvals_types,
            int8_t retval_count,
            char** err);

// Free loaded entity
void free_xcall(xcall* pxcall, char** err);
\end{minted}

The function \verb|load_runtime| loads $L$ runtime, or attaches to it if it already exists. For instance, in JVM, if MetaFFI is running from within a Java process, in some cases the MetaFFI runtime must be loaded (as in code listing \ref{code:py3-callable}). Therefore, \code{load\_runtime} initially checks if the runtime exists using \code{JNI\_GetCreatedJavaVMs}, if it exists, the plugin loads the existing runtime using \code{JNI\_GetCreatedJavaVMs}. In case a runtime does not exist, it creates a JVM runtime using \code{JNI\_CreateJavaVM}. In Python3.11, the plugin checks if Python3.11 is loaded using \code{Py\_IsInitialized}, and creates an interpreter using \code{Py\_InitializeEx}. A Python3.12 runtime plugin can support loading multiple interpreters for each runtime created \cite{PEP684}.  If $L$ has no runtime to load, like in C++, the function does not do anything.

The function \verb|free_runtime| releases the loaded runtime. For example, in JVM, the plugin calls the method \code{DestroyJavaVM} and in Python3.11, the function calls \code{Py\_FinalizeEx}. If there is no need to release the runtime (like in C++), or the runtime does not support releasing the runtime (like Go), the function does not do anything.

The function \verb|load_function| loads an entity from $L$. The implemented runtime plugin should define the required \emph{function path} that $L$ needs in order to load the entity, using the $key_i=value_i,tag_j$ notation. The plugin SDK provides the C++ class \code{function\_path\_parser} which parses the function path. The function loads the entity on the basis of the function path. For example, the JVM plugin uses JNI to load classes into \code{jclass}, methods into \code{jmethodID} or fields into \code{jfieldID}. The Python plug-in uses \code{PyObject\_GetAttrString} to find the requested entity and stores it in \code{PyObject*}.

The loaded entities (i.e. \code{jmethod} or \code{PyObject*}) required to call or use the entity, should be stored in a context structure that is returned to the \code{load\_entity} caller. This context structure is then passed to the XCall entry point, where it uses the context to call or use the entity.

The function returns an XCall, which is the array \verb|[void* XCall entrypoint, void* to context]|. The XCall entry point is discussed in section \ref{sec:l-xcall} . In case no context is required, simply pass \code{NULL}.

The function \verb|load_callable| receives an entity of $L$, constructs a context with entities required to call or use the target entity, and returns XCall similar to \code{load\_entity}. For example, in JVM, \code{load\_callable} receives \code{jobject} holding \code{Method} object. The function extracts \code{jclass} and \code{jmethodID} of the method wrapped within the code{Method} object and finally creates a context to be used when using the entity, returning \verb|[void* XCall entrypoint, void* to context]|.

The function \verb|free_function| frees the context related to the XCall and any entities inside it that require to be freed.

\subsubsection{Implementing XCall entry point} \label{sec:l-xcall}

The XCall entry point, as discussed in section \ref{sec:xcall}, contains one of the following signatures:
\begin{minted}[bgcolor=CodeBG,fontsize=\footnotesize,autogobble,tabsize=2,linenos]{c}
// For foreign entities with parameters or return values
void ()(void context, cdts* pcdts, char** out_err, uint64_t* out_err_len);

// For entities without parameters or return values
void ()(void context, char** out_err, uint64_t* out_err_len);
\end{minted}

When there is no need for a guest compiler, it is beneficial to integrate the entry points into the runtime plugin. This allows the XCall structure to return the relevant function pointer. For the JVM and Python3.11 plugins, the entry points were implemented as follows:
\begin{minted}[bgcolor=CodeBG,fontsize=\footnotesize,autogobble,tabsize=2,linenos]{c}
// Entry point for entities with parameters and return values
void xcall_params_ret(void* context, cdts params_ret[2], char** out_err, uint64_t* out_err_len);

// Entry point for entities with parameters but without return values
void xcall_params_no_ret(void* context, cdts parameters[2], char** out_err, uint64_t* out_err_len);

// Entry point for entities without parameters but with return values
void xcall_no_params_ret(void* context, cdts return_values[2], char** out_err, uint64_t* out_err_len);

// Entry point for entities without parameters and return values
void xcall_no_params_no_ret(void* context, char** out_err, uint64_t* out_err_len); 
\end{minted}

The \verb|xcall_params_ret| entry point is responsible for parsing the parameters CDTS located at \code{params\_ret[0]}. It then invokes the entity, utilizing the context if necessary, and fills in the return values in CDTS at \code{params\_ret[1]}. If an error occurs, an error message is written into memory that has been allocated on the heap using \code{malloc}. This error message is stored in the variable \code{out\_err}, and the length of the message is stored in the variable \code{out\_err\_len}.

XLLR exports C functions that can parse parameters from \code{params\_ret[0]} and write the return values to \code{params\_ret[1]}. To handle the manipulation of CDTs from CDTS, the plugin SDK provides the \code{load\_xllr\_api} function, which loads the exported functions of XLLR. Furthermore, the plugin SDK includes the \code{cdts\_wrapper} C++ class, which can be utilized for working with CDTS.

In the JVM and Python3.11 plugins, we extend \code{cdts\_wrapper}, and update its methods to align with the runtime. For instance, in JVM, to set \code{jvalue}, which is a union that holds any JVM entity, to CDT at index $i$ in the CDTS, one would invoke the \code{cdts\_java\_wrapper::from\_jvalue} method, passing the \code{jvalue}, its MetaFFI type, and index $i$. To retrieve a \code{jvalue} from CDT at index 
$i$ in the CDTS, one would invoke \code{cdts\_java\_wrapper::to\_jvalue}, passing $i$.

The remaining functions, \verb|xcall_no_params_ret|, \verb|xcall_params_no_ret| and \verb|xcall_no_params_no_ret|, operate similarly to \verb|xcall_params_ret|, but each performs only a portion of the process, as suggested by their names.

\subsubsection{Implementing Guest Compiler} \label{sec:l-guest-compiler}

Certain runtimes, such as Go, do not support the loading of their entities from their runtime executable code modules. As a result, MetaFFI requires a compiler to either \emph{generate an XCall entry point for each entity} or \emph{generate an entry point that the $L$ runtime can load}.

In order to implement the guest compiler, a dynamic library must be constructed that exports the following functions\footnote{In the future, guest compiler as a MetaFFI module will be supported}:

\begin{minted} [bgcolor=CodeBG,fontsize=\footnotesize,autogobble,tabsize=2,linenos]{c}
// This function is invoked once when the compiler plugin is loaded to perform any 
// necessary initialization tasks.
void init();

// This function is invoked to compile the L's entities,
//which are described by the IDL definition, to the guest executable module.
void compile_to_guest(const char* idl_def_json, uint32_t idl_def_json_length,
                                const char* output_path, uint32_t output_path_length,
                                const char* guest_options, uint32_t guest_options_length,
                                char** out_err, uint32_t* out_err_len); 
\end{minted}

The MetaFFI tool invokes the \verb|compile_to_guest| function to create the $L$ module with MetaFFI entry points. Initially, the MetaFFI tool calls $L$'s IDL plugin (refer to section \ref{sec:l-idl-plugin}) to generate the MetaFFI IDL, which describes the entities' signatures. Subsequently, it invokes \verb|compile_to_guest|, passing the IDL, the output path for writing the generated module, and the guest options obtained from the MetaFFI tool command line arguments.

As demonstrated in the implementation of a host compiler in section \ref{sec:l-guest-compiler}, the MetaFFI SDK provides Go code to generate the guest compiler dynamic library. The \code{NewCompilerPluginMain} function expects an instance of a host compiler and an instance of a guest compiler. In the absence of a host compiler implementation, the \code{Compile} method for the host compiler should be implemented, and a "Not Implemented" error should be returned.

The guest compiler is a Go struct that implements the \code{Compile} method:

\begin{minted}[bgcolor=CodeBG,fontsize=\footnotesize,autogobble,tabsize=2,linenos]{go}
type GuestCompiler struct {}
func (c *GuestCompiler) Compile(definition *IDL.IDLDefinition, 
                                outputDir string,
                                outputFilename string,
                                guestOptions map[string]string) (err error)
\end{minted}

This method provides the IDL JSON string as IDL Go entities (as detailed in \ref{sec:idl-details}). Additionally, any guest options, if present, have been parsed into a \code{map}. 

In the case of our implementation of the Go guest compiler plugin, the plugin generates Go source code, creating an entry point for each entity. The plugin employs the Go \code{text/template} package to implement code generation by creating repeated patterns in the $L$ source code based on the entities in the provided IDL. An implementer has the option to select from various languages to develop the plugin, although the MetaFFI SDK for compilers can be used with Go or any other languages supported by MetaFFI.

Each entry point is a C-exported function (utilizing CGo \cite{cgo_export_c}) with the expected XCall entry point signature. The plugin then compiles the source code using a Go command to construct a dynamic library for Go: \verb|go build -buildmode=c-shared -gcflags=-shared|.

During runtime, Go's runtime plugin loads the generated module and, using the received function path, loads the correct C-exported entry point. 

The Go guest compiler and guest template can be found in the official MetaFFI GitHub repository.

\subsection{$L$ IDL Plugin} \label{sec:l-idl-plugin}

The Interface Definition Language (IDL) plugin is designed to parse the source code or compiled modules of a given language $L$, and extract the signatures of the entities within. For callable entities, the plugin extracts the callable signature. For global variables and fields, it extracts a getter and a setter, depending on their accessibility. We recommend defining global scripts as functions and using an additional tag in the \emph{function path} to indicate that it is a global script. If $L$ provides access control, we also recommend extracting only \emph{public} entities.

The IDL plugin is a dynamic library that exports the following functions\footnote{In future iterations, the IDL plugin will have the capability to be a MetaFFI module}:
\begin{minted} [bgcolor=CodeBG,fontsize=\footnotesize,autogobble,tabsize=2,linenos]{c}
enum idl_input_type
{
    source_code = 0,
    path = 1
};

// generates MetaFFI IDL from a given source code, file or path
char* parse_idl(idl_input_type input_type,
						 const char* data,
						 char** out_err);
\end{minted}

The MetaFFI tool invokes the \verb|parse_idl| function to generate the MetaFFI IDL, which describes the signatures of the entities. The function receives the source code of $L$ or a path to the source code of $L$ or an executable module of the code runtime of $L$. The function should parse the source code or module and return the IDL as a string allocated with the XLLR allocation function. In case of an error, the error message is set to \code{out\_err} XLLR allocated string.

The MetaFFI SDK provides Go code to generate the IDL compiler dynamic library:
\begin{minted}[bgcolor=CodeBG,fontsize=\footnotesize,autogobble,tabsize=2,linenos]{go}
// PluginMain.go
package main

import "C"
import . "github.com/MetaFFI/plugin-sdk/compiler/go"

//export init_plugin
func init_plugin() { 
    CreateIDLPluginInterfaceHandler(NewGoIDLCompiler())
}
func main() {}
\end{minted}

The \code{CreateIDLPluginInterfaceHandler} function expects an instance of an IDL compiler that implements the method:
\begin{minted}[bgcolor=CodeBG,fontsize=\footnotesize,autogobble,tabsize=2,linenos]{go}
type IDLCompiler struct {}
func (c *IDLCompiler) ParseIDL(sourceCode string, filePath string) (*IDL.IDLDefinition, error)
\end{minted}

This method receives either the source code or the file path as passed to the \code{parse\_idl} C-function and returns an IDLDefinition object defined in the plugin SDK (detailed in section \ref{sec:idl-details}). As the method should parse the source code $L$ or the modules $L$, Go might not be a suitable language for that task and might not have available or up-to-date libraries to parse $L$ source code or modules. However, $L$ might provide a library to parse the syntax and modules of $L$ using the Abstract Syntax Tree (AST), reflection, or other libraries of $L$. Therefore, the IDL plugin can use MetaFFI to load the entities of $L$ to parse the code and modules of $L$. This process can be seen as a practical bootstrapping of the plugin. 

If $L$ requires a compiler plugin, to use $L$ entities from the IDL plugin, the implementer needs to manually write the MetaFFI JSON-based IDL for the required $L$ library, or use MetaFFI SDK IDL entities to generate the IDL (as detailed in section \ref{sec:idl-details}). Once the implementer has a MetaFFI IDL, it can be passed to the MetaFFI tool as IDL: \verb|metaffi -c --idl [JSON file] -g|

As an example, for the Python3.11 IDL plugin, we have written a \code{py\_extractor} Python class that uses Python's \code{inspect} package, which extracts signatures of Python entities. To use this class, we are using MetaFFI to load the class into Go. Similarly, in the JVM IDL plugin, we have implemented a \code{JavaExtractor} class which uses the \code{asm} package to parse \code{.class} and \code{.jar} files and the \code{javaparser} package to parse \code{.java} files.

The source code for the Python3.11 IDL and JVM IDL plugins is available in the MetaFFI public GitHub repository.

\section[IDL Details]{IDL Details \technical} \label{sec:idl-details}

\begin{figure}[h] 
\centering
\includegraphics[width=.5\linewidth]{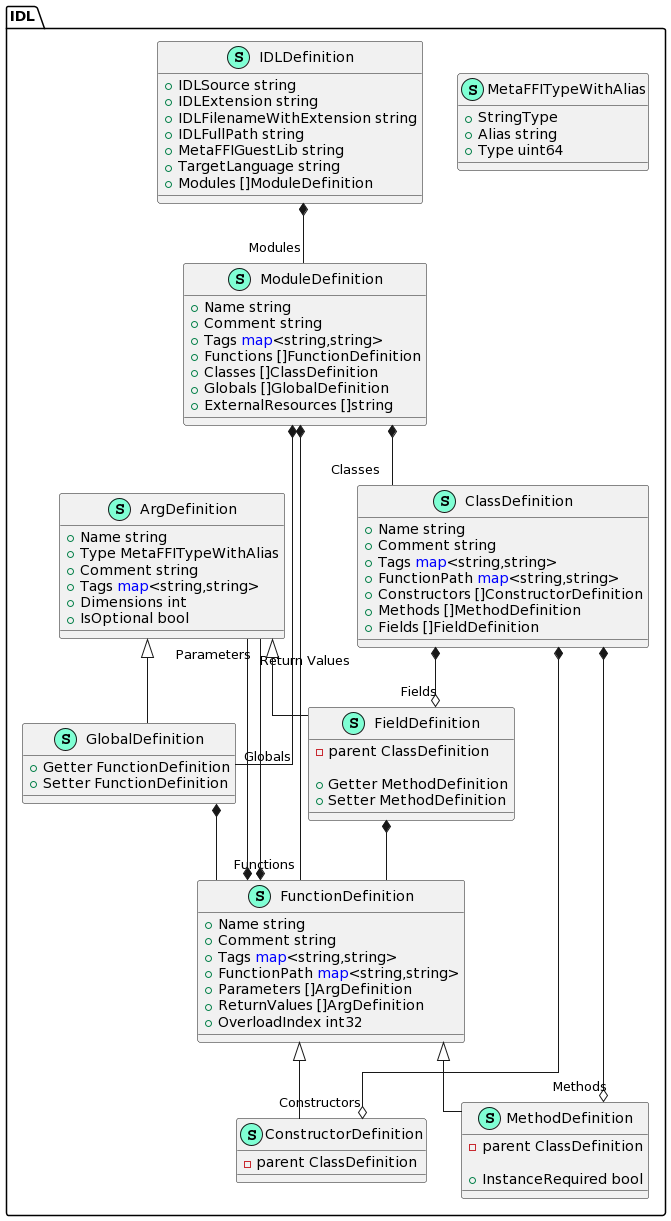}
\caption{UML of MetaFFI IDL Entities}
\label{fig:idl-uml}
\end{figure}

The MetaFFI Interface Definition Language (IDL) is a hierarchical structure that can be represented as a JSON-formatted string, as shown in Figure \ref{fig:simplified-def-model}. The MetaFFI SDK offers Go entities to construct the IDL, with its root being the \code{IDLDefinition} (Figure \ref{fig:idl-uml}).

The SDK provides the function \code{NewIDLDefinitionFromJSON} to load a \code{IDLDefinition} from a given IDL, and the methods \code{IDLDefinition.FinalizeConstruction} and \code{IDLDefinition.ToJSON} to generate an IDL from the object \code{IDLDefinition}. It is important to note that these Go entities can be utilized in any language supported by MetaFFI.

This section provides a detailed explanation of the fields and their roles in the IDL.

There are several common fields in the IDL entities:
\begin{itemize}
    \item \code{Name}: This field holds the name of the entity.
    \item \code{Comment}: This field holds comments about the entity. These are usually extracted from the source and can be placed within the generated wrappers of a host compiler to provide information and documentation about the entity.
    \item \code{Tags}: This is a string-to-string map, which allows plugins to extend the entity with additional, plugin-specific fields.
\end{itemize}

\subsection{GlobalDefinition}

The root of the IDL, \code{GlobalDefinition}, holds information about the IDL modules it contains. Its fields are as follows:

\begin{itemize}
    \item \code{IDLSource}: This field holds the file or path from which the IDL was extracted.
    \item \code{IDLExtension}: This field holds the extension of the source file, if applicable.
    \item \code{IDLFileNameWithExtension}: This field holds the name of the source with its extension, if applicable.
    \item \code{IDLFullPath}: This field holds the full path of the source.
    \item \code{MetaFFIGuestLib}: This field holds the name of the generated module.
    \item \code{Modules}: This field holds a collection of \code{ModuleDefinition}, which represent modules defined in the IDL and points to the rest of the IDL tree. 
\end{itemize}

\subsection{ModuleDefinition}

The \code{ModuleDefinition} represents a module, which is a collection of globals, functions, and complex types. Its fields are as follows:

\begin{itemize}
    \item \textbf{Name}
    \item \textbf{Comment}
    \item \textbf{Tags}
    \item \textbf{IDLFullPath}: This field holds the full path of the source.
    \item \textbf{Functions}: This field holds a collection of \code{FunctionDefinition}, which represent functions defined in the module.
    \item \textbf{Classes}: This field holds a collection of \code{ClassDefinition}, which represent complex types defined in the module.
    \item \textbf{Globals}: This field holds a collection of \code{ArgDefinition}, which represent global variables or constants defined in the module.
    \item \textbf{ExternalResources}: This field is an array of strings containing external resources a guest compiler might require to perform its task. For example, dependent modules or configuration files. \end{itemize}

\subsection{MetaFFITypeInfo}

The \code{MetaFFITypeInfo} holds the MetaFFI Type in its string form for easier readability and in its numeric form. It also holds an alias for the type if such an alias exists. In addition, it stores the dimensions of the type, where a value greater than zero means that the type is an array. Both the string and the numeric value of the MetaFFI types are defined in \code{metaffi\_primitives.h} in the SDK. For the simple usage of IDL entities in Go, it is also available in Go in \code{MetaFFIPrimitives.go}.

\subsection{ArgDefinition}

The \code{ArgDefinition} represents a variable which can be a global, field, argument, or return value. Its fields are as follows:

\begin{itemize}
    \item \code{Name}
    \item \code{Type}: This field is the MetaFFI Type of the variable and defined by \code{MetaFFITypeInfo}.
    \item \code{Comment}
    \item \code{Tags}
    \item \code{Dimensions}: This field holds the number of dimensions of the variable. If $Dimensions$ is 0, the variable doesn’t hold an array. If it is larger than 0, it is an array with $Dimensions$ dimensions.
    \item \code{IsOptional}: This field holds if the argument is optional or not. This is used in case \code{ArgDefinition} is used as an argument.
\end{itemize}

\subsection{FunctionDefinition}

The \code{FunctionDefinition} represents a function or a callable that has no link to a complex structure. Its fields are as follows:

\begin{itemize}
    \item \code{Name}
    \item \code{Comment}
    \item \code{Tags}
    \item \code{FunctionPath}: This field holds the function path required to load the function. It is parsed into a string-to-string map. The mapped-value of Tags in \code{FunctionPath} should be ignored.
    \item \code{Parameters}: This field contains an array of \code{ArgDefinition}, representing the parameters the function expects.
    \item \code{ReturnValues}: This field contains an array of \code{ArgDefinition}, representing the return values the function returns.
    \item \code{OverloadIndex}: This field is a numeric number used to differentiate between multiple functions with the same name, as might occur in languages supporting overloaded functions and methods. \end{itemize}

\subsection{GlobalDefinition}

The \code{GlobalDefinition} represents a global variable which, like a function, does not have a link to a complex structure. \code{GlobalDefinition} extends \code{ArgDefinition} and adds the following fields:

\begin{itemize}
    \item \code{Getter}: This field holds a \code{FunctionDefinition} that has no parameters and a return value of the \code{GlobalDefinition} Type. If the global is write-only, Getter is null.
    \item \code{Setter}: This field holds a \code{FunctionDefinition} that accepts the \code{GlobalDefinition} Type and does not return anything. If the global is read-only, Setter is null. \end{itemize}

\subsection{ClassDefinition} 

The \code{ClassDefinition} represents a complex type with the added features of callables linked to the complex type (i.e., methods). Its fields are as follows:

\begin{itemize}
    \item \code{FunctionPath}: This field holds the part of the function path required to load the entities within the class. The method \code{IDLDefinition.FinalizeConstruction} uses the FunctionPath in \code{ClassDefinition} to construct the function path of the entities in the class. Notice, this field is not required in the JSON IDL, and used a helper field.
    \item \code{Constructors}: This field holds a collection of \code{ConstructorDefinition}.
    \item \code{Methods}: This field holds a collection of \code{MethodDefinition}.
    \item \code{Fields}: This field holds a collection of \code{FieldDefinition}.
\end{itemize}

\subsection{ConstructorDefinition}

The \code{ConstructorDefinition} represents a function that creates a complex type. In object-oriented languages, \code{ConstructorDefinition} represents a constructor, in others it can represent a function returning an instance of the \code{ClassDefinition} type. \code{ConstructorDefinition} extends \code{FunctionDefinition}, and it returns, at least, an instance of the \code{ClassDefinition} type. It also has an additional \code{parent} field, pointing to the \code{ClassDefinition} entity that holds this \code{ConstructorDefinition}.

\subsection{MethodDefinition}

The \code{MethodDefinition} represents a function linked to a complex type. \code{MethodDefinition} extends \code{FunctionDefinition} but adds two additional fields:

\begin{itemize}
    \item \code{parent}: This field holds a pointer to the \code{ClassDefinition} entity holding this \code{MethodDefinition}.
    \item \code{InstanceRequired}: This field is a boolean which states if the function requires an instance of the \code{ClassDefinition}. For example, in Java, a static method is part of the class, but does not require an instance of the class. We are aware that it is possible to look at static methods as functions, but we do not want to break this link in the IDL. In any case, one can define a static method as a function and provide the correct function path. If an instance is required, then the method expects an instance of the \code{ClassDefinition} type as its first parameter. 
\end{itemize}

\subsection{FieldDefinition}

The \code{FieldDefinition} is a representation of a field within a complex type. 

Similar to \code{GlobalDefinition}, \code{FieldDefinition} is an extension of \code{ArgDefinition}. However, it differs from \code{GlobalDefinition} in that \code{Getter} and \code{Setter} are \code{MethodDefinitions}, as opposed to \code{FunctionDefinitions}. 

In addition, \code{FieldDefinition} contains a \code{parent} field, similar to \textit{MethodDefinition}. The \textit{InstanceRequired} field within the \textit{MethodDefinitions} specifies whether or not the field requires an instance (e.g. Java static field).

\section[Future Work]{Future Work \academic}
\label{sec:future_work}

Further research is still needed on the MetaFFI system. Additionally, the method of loading and overseeing multiple runtimes brings forth new inquiries. The subsequent questions necessitate additional investigation:

\begin{itemize}

\item \large{\textbf{MetaFFI is the first \emph{simple interoperability tool}\cite{empirical_multi_lingual}, is it really easier to use by a single developer? by a team collaborating code?}}\\
MetaFFI is a \emph{simple interoperability tool} as defined in \cite{empirical_multi_lingual}, but is it really easy to use? In the experiment, we will pick volunteers who are familiar with at least 2 of the MetaFFI supported languages. We will make sure that the distribution of the picked languages is similar between all languages. Volunteers will perform specific tasks and rate their experience with MetaFFI. 

\item \large{\textbf{What is the impact on the performance of MetaFFI indirect interoperability compared to direct interoperability? Moreover, call within the same language?}} Evaluate the performance of xcall by comparing it to other interoperability mechanisms and direct calls within the same language. This is to evaluate the performance hit of xcall.

\item \large{\textbf{How to debug in multiple runtimes without handling multiple debuggers?}}
Developing in a multilingual environment is not just writing code but also debugging it. That means loading and using several debuggers. Some debuggers cannot be attached to the same process at the same time. Research to improve debuggers to work together is essential to allow developers to debug multiple languages and not just use them.

%

\end{itemize}

\section{Conclusion} \label{sec:conclusions}
MetaFFI is the first mechanism to provide\emph{simple interoperability} presented in the Introduction and detailed in \cite{empirical_multi_lingual}. The system offers loading of foreign modules and foreign entities in a similar concept to loading a C/C++ dynamic libraries exported function and classes. With the usage of Common Data Types, MetaFFI provides interoperability across a wide range of programming languages using a uniform API. Moreover, using a compiler, MetaFFI can provide a completely transparent interoperability experience to the user.

MetaFFI utilizes shallow binding mechanisms and the widespread use of C binding \cite{empirical_multi_lingual} to offer extensive binding capabilities, enabling the access of external entities such as functions, objects, fields, and so on. By managing multiple runtimes in the same process, MetaFFI allows the use of languages in their designated runtimes. Using Common Data Types (CDTs), MetaFFI offers a means to transfer parameters and results across languages without requiring distinct intermediary solutions for each language pair. By opting for this indirect strategy that utilizes interoperability mechanisms with C and incorporates CDTs, MetaFFI requires $n$ mechanisms for interoperability between $n$ languages, in contrast to the $n^2$ mechanisms needed by direct methods. MetaFFI automatically generates the IDL to improve its transparency in usage, eliminating the need for manual creation and reducing the associated workload \cite{polispin}. MetaFFI also offers the capability to pass complex types and callback functions by reference, enabling developers to interact with foreign entities as if they were part of their own programming language.

The features of MetaFFI enable users to work with other languages seamlessly within the familiar environment of the host language, without the need to switch contexts. The indirect method of achieving interoperability through the management of various runtime environments and CDTs, meeting the requirements of \emph{simple interoperability}\cite{empirical_multi_lingual}, demonstrates a notable enhancement in the quantity of necessary interoperability mechanisms and offers a straightforward and user-friendly approach for multilingual programming.

\bibliography{main}


\begin{thebibliography}{58}


\ifx \showCODEN    \undefined \def \showCODEN     #1{\unskip}     \fi
\ifx \showDOI      \undefined \def \showDOI       #1{#1}\fi
\ifx \showISBNx    \undefined \def \showISBNx     #1{\unskip}     \fi
\ifx \showISBNxiii \undefined \def \showISBNxiii  #1{\unskip}     \fi
\ifx \showISSN     \undefined \def \showISSN      #1{\unskip}     \fi
\ifx \showLCCN     \undefined \def \showLCCN      #1{\unskip}     \fi
\ifx \shownote     \undefined \def \shownote      #1{#1}          \fi
\ifx \showarticletitle \undefined \def \showarticletitle #1{#1}   \fi
\ifx \showURL      \undefined \def \showURL       {\relax}        \fi
\providecommand\bibfield[2]{#2}
\providecommand\bibinfo[2]{#2}
\providecommand\natexlab[1]{#1}
\providecommand\showeprint[2][]{arXiv:#2}

\bibitem[dlo({[n.\,d.]})]%
        {dlopen_die}
 \bibinfo{year}{[n.\,d.]}\natexlab{}.
\newblock \bibinfo{title}{dlopen(3) - {Linux} man page}.
\newblock
\newblock
\urldef\tempurl%
\url{https://linux.die.net/man/3/dlopen}
\showURL{%
\tempurl}


\bibitem[dls({[n.\,d.]})]%
        {dlsym_die}
 \bibinfo{year}{[n.\,d.]}\natexlab{}.
\newblock \bibinfo{title}{dlsym(3) - {Linux} man page}.
\newblock
\newblock
\urldef\tempurl%
\url{https://linux.die.net/man/3/dlsym}
\showURL{%
\tempurl}


\bibitem[cla({[n.\,d.]})]%
        {classpath}
 \bibinfo{year}{[n.\,d.]}\natexlab{}.
\newblock \bibinfo{title}{{PATH} and {CLASSPATH} ({The} {Java}™ {Tutorials} {\textgreater} {Essential} {Java} {Classes} {\textgreater} {The} {Platform} {Environment})}.
\newblock
\newblock
\urldef\tempurl%
\url{https://docs.oracle.com/javase/tutorial/essential/environment/paths.html}
\showURL{%
\tempurl}


\bibitem[PEP({[n.\,d.]})]%
        {PEP684}
 \bibinfo{year}{[n.\,d.]}\natexlab{}.
\newblock \bibinfo{title}{{PEP} 684 – {A} {Per}-{Interpreter} {GIL} {\textbar} peps.python.org}.
\newblock
\newblock
\urldef\tempurl%
\url{https://peps.python.org/pep-0684/}
\showURL{%
\tempurl}


\bibitem[Beazley and {others}(1996)]%
        {swig}
\bibfield{author}{\bibinfo{person}{David~M Beazley} {and} \bibinfo{person}{{others}}.} \bibinfo{year}{1996}\natexlab{}.
\newblock \showarticletitle{{SWIG}: {An} easy to use tool for integrating scripting languages with {C} and {C}++.}. In \bibinfo{booktitle}{\emph{Tcl/{Tk} workshop}}, Vol.~\bibinfo{volume}{43}.
\newblock


\bibitem[Berners-Lee and Connolly(1995)]%
        {html_rfc}
\bibfield{author}{\bibinfo{person}{Tim Berners-Lee} {and} \bibinfo{person}{Daniel~W. Connolly}.} \bibinfo{year}{1995}\natexlab{}.
\newblock \bibinfo{title}{Hypertext markup language - 2.0}.
\newblock
\newblock
\urldef\tempurl%
\url{https://doi.org/10.17487/RFC1866}
\showDOI{\tempurl}
\newblock
\shownote{Number: 1866 Series: Request for comments tex.howpublished: RFC 1866 tex.pagetotal: 77}.


\bibitem[Bray(2017)]%
        {json_rfc}
\bibfield{author}{\bibinfo{person}{Tim Bray}.} \bibinfo{year}{2017}\natexlab{}.
\newblock \bibinfo{title}{The {JavaScript} object notation ({JSON}) data interchange format}.
\newblock
\newblock
\urldef\tempurl%
\url{https://doi.org/10.17487/RFC8259}
\showDOI{\tempurl}
\newblock
\shownote{Number: 8259 Series: Request for comments tex.howpublished: RFC 8259 tex.pagetotal: 16}.


\bibitem[{Charles Oliver Nutter} et~al\mbox{.}(2022)]%
        {jruby}
\bibfield{author}{\bibinfo{person}{{Charles Oliver Nutter}}, \bibinfo{person}{{Thomas Enebo}}, \bibinfo{person}{{Ola Bini}}, {and} \bibinfo{person}{{Nick Sieger}}.} \bibinfo{year}{2022}\natexlab{}.
\newblock \bibinfo{title}{{JRuby} - an implementation of the {Ruby} language on the {JVM}}.
\newblock
\newblock
\urldef\tempurl%
\url{https://github.com/jruby/jruby}
\showURL{%
\tempurl}


\bibitem[Cherny-Shahar and Yehudai(2024)]%
        {empirical_multi_lingual}
\bibfield{author}{\bibinfo{person}{Tsvi Cherny-Shahar} {and} \bibinfo{person}{Amiram Yehudai}.} \bibinfo{year}{2024}\natexlab{}.
\newblock \bibinfo{title}{Empirical {Study} on {Multi}-{Lingual} \& {Interoperability} {Usage} [{In} preparation]}.
\newblock
\newblock


\bibitem[Chiba(2019)]%
        {code_migration}
\bibfield{author}{\bibinfo{person}{Shigeru Chiba}.} \bibinfo{year}{2019}\natexlab{}.
\newblock \showarticletitle{Foreign language interfaces by code migration}. In \bibinfo{booktitle}{\emph{Proceedings of the 18th {ACM} {SIGPLAN} international conference on generative programming: {Concepts} and experiences}} \emph{(\bibinfo{series}{{GPCE} 2019})}. \bibinfo{publisher}{Association for Computing Machinery}, \bibinfo{address}{New York, NY, USA}, \bibinfo{pages}{1--13}.
\newblock
\showISBNx{978-1-4503-6980-0}
\urldef\tempurl%
\url{https://doi.org/10.1145/3357765.3359521}
\showDOI{\tempurl}
\newblock
\shownote{Number of pages: 13 Place: Athens, Greece}.


\bibitem[Chisnall(2013)]%
        {challenges_of_interop}
\bibfield{author}{\bibinfo{person}{David Chisnall}.} \bibinfo{year}{2013}\natexlab{}.
\newblock \showarticletitle{The challenge of cross-language interoperability}.
\newblock \bibinfo{journal}{\emph{Queue}} \bibinfo{volume}{11}, \bibinfo{number}{10} (\bibinfo{date}{Oct.} \bibinfo{year}{2013}), \bibinfo{pages}{20--28}.
\newblock
\showISSN{1542-7730}
\urldef\tempurl%
\url{https://doi.org/10.1145/2542661.2543971}
\showDOI{\tempurl}
\newblock
\shownote{Number of pages: 9 Place: New York, NY, USA Publisher: Association for Computing Machinery tex.issue\_date: October 2013}.


\bibitem[Contributors(2021)]%
        {gobject_wiki}
\bibfield{author}{\bibinfo{person}{Wikipedia Contributors}.} \bibinfo{year}{2021}\natexlab{}.
\newblock \bibinfo{title}{{GObject}}.
\newblock
\newblock
\urldef\tempurl%
\url{https://en.wikipedia.org/wiki/GObject}
\showURL{%
\tempurl}


\bibitem[contributors(2022)]%
        {idl_wikipedia}
\bibfield{author}{\bibinfo{person}{Wikipedia contributors}.} \bibinfo{year}{2022}\natexlab{}.
\newblock \bibinfo{title}{Interface description language — {Wikipedia}, the free encyclopedia}.
\newblock
\newblock
\urldef\tempurl%
\url{https://en.wikipedia.org/w/index.php?title=Interface_description_language&oldid=1064057807}
\showURL{%
\tempurl}


\bibitem[Corporation(2020)]%
        {ms_dotnet}
\bibfield{author}{\bibinfo{person}{Microsoft Corporation}.} \bibinfo{year}{2020}\natexlab{}.
\newblock \bibinfo{title}{.{NET} {\textbar} free. {Cross}-platform. {Open} source. 2020}.
\newblock
\newblock
\urldef\tempurl%
\url{https://dotnet.microsoft.com/}
\showURL{%
\tempurl}


\bibitem[Cowell(1996)]%
        {ole}
\bibfield{author}{\bibinfo{person}{John Cowell}.} \bibinfo{year}{1996}\natexlab{}.
\newblock \showarticletitle{Object linking and embedding ({OLE})}.
\newblock In \bibinfo{booktitle}{\emph{Essential visual basic 4.0 fast: {How} to develop applications in visual basic}}. \bibinfo{publisher}{Springer London}, \bibinfo{address}{London}, \bibinfo{pages}{136--144}.
\newblock
\showISBNx{978-1-4471-3093-2}
\urldef\tempurl%
\url{https://doi.org/10.1007/978-1-4471-3093-2_18}
\showDOI{\tempurl}


\bibitem[Crowl(2006)]%
        {ThreadLocalStorage}
\bibfield{author}{\bibinfo{person}{Lawrence Crowl}.} \bibinfo{year}{2006}\natexlab{}.
\newblock \bibinfo{title}{Thread-{Local} {Storage}}.
\newblock
\newblock
\urldef\tempurl%
\url{https://www.open-std.org/jtc1/sc22/wg21/docs/papers/2006/n1966.html}
\showURL{%
\tempurl}


\bibitem[{David M. Beazley}(2008)]%
        {swig_master_class}
\bibfield{author}{\bibinfo{person}{{David M. Beazley}}.} \bibinfo{year}{2008}\natexlab{}.
\newblock \bibinfo{title}{{SWIG} {Master} {Class}}.
\newblock
\newblock


\bibitem[Flanagan(2016)]%
        {css_rfc}
\bibfield{author}{\bibinfo{person}{Heather Flanagan}.} \bibinfo{year}{2016}\natexlab{}.
\newblock \bibinfo{title}{Cascading style sheets ({CSS}) requirements for {RFCs}}.
\newblock
\newblock
\urldef\tempurl%
\url{https://doi.org/10.17487/RFC7993}
\showDOI{\tempurl}
\newblock
\shownote{Number: 7993 Series: Request for comments tex.howpublished: RFC 7993 tex.pagetotal: 14}.


\bibitem[Foundation(2020a)]%
        {log4net}
\bibfield{author}{\bibinfo{person}{Apache~Software Foundation}.} \bibinfo{year}{2020}\natexlab{a}.
\newblock \bibinfo{title}{Apache log4net: {A} .{NET} framework logging library}.
\newblock
\newblock
\urldef\tempurl%
\url{https://logging.apache.org/log4net/index.html}
\showURL{%
\tempurl}


\bibitem[Foundation(2021a)]%
        {log4j2}
\bibfield{author}{\bibinfo{person}{Apache~Software Foundation}.} \bibinfo{year}{2021}\natexlab{a}.
\newblock \bibinfo{title}{Apache log4j 2.14.1}.
\newblock
\newblock
\urldef\tempurl%
\url{https://logging.apache.org/log4j/2.x/}
\showURL{%
\tempurl}


\bibitem[Foundation(2024)]%
        {log4cxx}
\bibfield{author}{\bibinfo{person}{Apache~Software Foundation}.} \bibinfo{year}{2024}\natexlab{}.
\newblock \bibinfo{title}{Apache log4cxx: {A} logging framework for {C}++}.
\newblock
\newblock
\urldef\tempurl%
\url{https://logging.apache.org/log4cxx/latest_stable/index.html}
\showURL{%
\tempurl}


\bibitem[Foundation(2021b)]%
        {haxe}
\bibfield{author}{\bibinfo{person}{Haxe Foundation}.} \bibinfo{year}{2021}\natexlab{b}.
\newblock \bibinfo{title}{Haxe - {The} {Cross}-platform toolkit}.
\newblock
\newblock
\urldef\tempurl%
\url{https://haxe.org/}
\showURL{%
\tempurl}


\bibitem[Foundation(2022)]%
        {ironpython}
\bibfield{author}{\bibinfo{person}{.NET Foundation}.} \bibinfo{year}{2022}\natexlab{}.
\newblock \bibinfo{title}{{IronPython}.net}.
\newblock
\newblock
\urldef\tempurl%
\url{https://ironpython.net/}
\showURL{%
\tempurl}


\bibitem[Foundation({[n.\,d.]})]%
        {cpython_garbage_collection}
\bibfield{author}{\bibinfo{person}{Python~Software Foundation}.} \bibinfo{year}{[n.\,d.]}\natexlab{}.
\newblock \bibinfo{title}{Design of {CPython}’s {Garbage} {Collector} - {Python} {Developer}'s {Guide}}.
\newblock
\newblock
\urldef\tempurl%
\url{https://devguide.python.org/garbage_collector/}
\showURL{%
\tempurl}


\bibitem[Foundation(2020b)]%
        {python_ctypes}
\bibfield{author}{\bibinfo{person}{Python~Software Foundation}.} \bibinfo{year}{2020}\natexlab{b}.
\newblock \bibinfo{title}{ctypes — {A} foreign function library for {Python} — {Python} 3.8.3 documentation}.
\newblock
\newblock
\urldef\tempurl%
\url{https://docs.python.org/3/library/ctypes.html}
\showURL{%
\tempurl}


\bibitem[Foundation(2021c)]%
        {jython}
\bibfield{author}{\bibinfo{person}{Python~Software Foundation}.} \bibinfo{year}{2021}\natexlab{c}.
\newblock \bibinfo{title}{Jython}.
\newblock
\newblock
\urldef\tempurl%
\url{https://www.jython.org/}
\showURL{%
\tempurl}


\bibitem[Gangemi(2021)]%
        {go_memory_management}
\bibfield{author}{\bibinfo{person}{Scott Gangemi}.} \bibinfo{year}{2021}\natexlab{}.
\newblock \bibinfo{title}{An overview of memory management in {Go}}.
\newblock
\newblock
\urldef\tempurl%
\url{https://medium.com/safetycultureengineering/an-overview-of-memory-management-in-go-9a72ec7c76a8}
\showURL{%
\tempurl}


\bibitem[{gewarren}(2019)]%
        {crl_overview}
\bibfield{author}{\bibinfo{person}{{gewarren}}.} \bibinfo{year}{2019}\natexlab{}.
\newblock \bibinfo{title}{Common language runtime ({CLR}) overview - .{NET} framework}.
\newblock
\newblock
\urldef\tempurl%
\url{https://docs.microsoft.com/en-us/dotnet/standard/clr?redirectedfrom=MSDN}
\showURL{%
\tempurl}


\bibitem[Green(2019)]%
        {libffi}
\bibfield{author}{\bibinfo{person}{Anthony Green}.} \bibinfo{year}{2019}\natexlab{}.
\newblock \bibinfo{title}{{LibFFI}}.
\newblock
\newblock
\urldef\tempurl%
\url{https://sourceware.org/libffi/}
\showURL{%
\tempurl}


\bibitem[Grimmer et~al\mbox{.}(2018)]%
        {trufflevm}
\bibfield{author}{\bibinfo{person}{Matthias Grimmer}, \bibinfo{person}{Roland Schatz}, \bibinfo{person}{Chris Seaton}, \bibinfo{person}{Thomas Würthinger}, \bibinfo{person}{Mikel Luján}, {and} \bibinfo{person}{Hanspeter Mössenböck}.} \bibinfo{year}{2018}\natexlab{}.
\newblock \showarticletitle{Cross-language interoperability in a multi-language runtime}.
\newblock \bibinfo{journal}{\emph{ACM Transactions on Programming Languages and Systems}} \bibinfo{volume}{40}, \bibinfo{number}{2} (\bibinfo{date}{May} \bibinfo{year}{2018}).
\newblock
\showISSN{0164-0925}
\urldef\tempurl%
\url{https://doi.org/10.1145/3201898}
\showDOI{\tempurl}
\newblock
\shownote{Number of pages: 43 Place: New York, NY, USA Publisher: Association for Computing Machinery tex.articleno: Article 8 tex.issue\_date: June 2018}.


\bibitem[Group(2020)]%
        {corba}
\bibfield{author}{\bibinfo{person}{Object~Management Group}.} \bibinfo{year}{2020}\natexlab{}.
\newblock \bibinfo{title}{About the common object request broker architecture specification version 3.4 beta}.
\newblock
\newblock
\urldef\tempurl%
\url{https://www.omg.org/spec/CORBA/}
\showURL{%
\tempurl}


\bibitem[International(2012)]%
        {ms_cil}
\bibfield{author}{\bibinfo{person}{Ecma International}.} \bibinfo{year}{2012}\natexlab{}.
\newblock \bibinfo{title}{Standard {ECMA}-335 2012}.
\newblock
\newblock
\urldef\tempurl%
\url{http://www.ecma-international.org/publications/standards/Ecma-335.htm}
\showURL{%
\tempurl}


\bibitem[Kaplan et~al\mbox{.}(1998)]%
        {polispin}
\bibfield{author}{\bibinfo{person}{A. Kaplan}, \bibinfo{person}{J. Ridgway}, {and} \bibinfo{person}{J.~C. Wileden}.} \bibinfo{year}{1998}\natexlab{}.
\newblock \showarticletitle{Why {IDLs} are not ideal}. In \bibinfo{booktitle}{\emph{Proceedings of the 9th international workshop on software specification and design}} \emph{(\bibinfo{series}{{IWSSD} '98})}. \bibinfo{publisher}{IEEE Computer Society}, \bibinfo{address}{USA}, \bibinfo{pages}{2}.
\newblock
\showISBNx{0-8186-8439-9}


\bibitem[karl-bridge microsoft(2022)]%
        {getprocaddress_msdn}
\bibfield{author}{\bibinfo{person}{karl-bridge microsoft}.} \bibinfo{year}{2022}\natexlab{}.
\newblock \bibinfo{title}{{GetProcAddress} function (libloaderapi.h) - {Win32} apps}.
\newblock
\newblock
\urldef\tempurl%
\url{https://learn.microsoft.com/en-us/windows/win32/api/libloaderapi/nf-libloaderapi-getprocaddress}
\showURL{%
\tempurl}


\bibitem[karl-bridge microsoft(2023)]%
        {loadlibrary_msdn}
\bibfield{author}{\bibinfo{person}{karl-bridge microsoft}.} \bibinfo{year}{2023}\natexlab{}.
\newblock \bibinfo{title}{{LoadLibraryA} function (libloaderapi.h) - {Win32} apps}.
\newblock
\newblock
\urldef\tempurl%
\url{https://learn.microsoft.com/en-us/windows/win32/api/libloaderapi/nf-libloaderapi-loadlibrarya}
\showURL{%
\tempurl}


\bibitem[Lattner and Adve(2004)]%
        {llvm}
\bibfield{author}{\bibinfo{person}{C. Lattner} {and} \bibinfo{person}{V. Adve}.} \bibinfo{year}{2004}\natexlab{}.
\newblock \showarticletitle{{LLVM}: a compilation framework for lifelong program analysis amp; transformation}. In \bibinfo{booktitle}{\emph{International symposium on code generation and optimization, 2004. {CGO} 2004.}} \bibinfo{pages}{75--86}.
\newblock
\urldef\tempurl%
\url{https://doi.org/10.1109/CGO.2004.1281665}
\showDOI{\tempurl}


\bibitem[Lindholm et~al\mbox{.}(2013)]%
        {jvm_specs}
\bibfield{author}{\bibinfo{person}{Tim Lindholm}, \bibinfo{person}{Frank Yellin}, \bibinfo{person}{Gilad Bracha}, {and} \bibinfo{person}{Alex Buckley}.} \bibinfo{year}{2013}\natexlab{}.
\newblock \bibinfo{title}{The java virtual machine specification}.
\newblock
\newblock
\urldef\tempurl%
\url{https://docs.oracle.com/javase/specs/jvms/se7/html/}
\showURL{%
\tempurl}


\bibitem[LLC({[n.\,d.]})]%
        {cgo_export_c}
\bibfield{author}{\bibinfo{person}{Google LLC}.} \bibinfo{year}{[n.\,d.]}\natexlab{}.
\newblock \bibinfo{title}{cgo - the go programming language 2020}.
\newblock
\newblock
\urldef\tempurl%
\url{https://golang.org/cmd/cgo/#hdr-C_references_to_Go}
\showURL{%
\tempurl}


\bibitem[Mayer et~al\mbox{.}(2017)]%
        {xlang_survey}
\bibfield{author}{\bibinfo{person}{Philip Mayer}, \bibinfo{person}{Michael Kirsch}, {and} \bibinfo{person}{Minh-Anh Le}.} \bibinfo{year}{2017}\natexlab{}.
\newblock \showarticletitle{On multi-language software development, cross-language links and accompanying tools: a survey of professional software developers}.
\newblock \bibinfo{journal}{\emph{Journal of Software Engineering Research and Development}}  \bibinfo{volume}{5} (\bibinfo{date}{Dec.} \bibinfo{year}{2017}).
\newblock
\urldef\tempurl%
\url{https://doi.org/10.1186/s40411-017-0035-z}
\showDOI{\tempurl}


\bibitem[{mcleanbyron}(2018)]%
        {com_component}
\bibfield{author}{\bibinfo{person}{{mcleanbyron}}.} \bibinfo{year}{2018}\natexlab{}.
\newblock \bibinfo{title}{Component object model ({COM}) - win32 apps}.
\newblock
\newblock
\urldef\tempurl%
\url{https://docs.microsoft.com/en-us/windows/win32/com/component-object-model--com--portal}
\showURL{%
\tempurl}


\bibitem[{Microsoft}(2018)]%
        {com_variant}
\bibfield{author}{\bibinfo{person}{{Microsoft}}.} \bibinfo{year}{2018}\natexlab{}.
\newblock \bibinfo{title}{{VARIANT} (oaidl.h) - win32 apps {\textbar} microsoft docs}.
\newblock
\newblock
\urldef\tempurl%
\url{https://docs.microsoft.com/en-us/windows/win32/api/oaidl/ns-oaidl-variant}
\showURL{%
\tempurl}


\bibitem[Microsystems({[n.\,d.]})]%
        {jvm_garbage_collector}
\bibfield{author}{\bibinfo{person}{Sun Microsystems}.} \bibinfo{year}{[n.\,d.]}\natexlab{}.
\newblock \showarticletitle{Memory {Management} in the {Java} {HotSpot} {Virtual} {Machine}}.
\newblock  (\bibinfo{year}{[n.\,d.]}).
\newblock


\bibitem[Nelson(1981)]%
        {rpc}
\bibfield{author}{\bibinfo{person}{Bruce~Jay Nelson}.} \bibinfo{year}{1981}\natexlab{}.
\newblock \showarticletitle{Remote procedure call}.
\newblock
\urldef\tempurl%
\url{https://api.semanticscholar.org/CorpusID:59728511}
\showURL{%
\tempurl}


\bibitem[Petersen({[n.\,d.]})]%
        {emacs-lisp-26-1}
\bibfield{author}{\bibinfo{person}{Mickey Petersen}.} \bibinfo{year}{[n.\,d.]}\natexlab{}.
\newblock \bibinfo{title}{What's {New} in {Emacs} 26.1}.
\newblock
\newblock
\urldef\tempurl%
\url{https://www.masteringemacs.org/article/whats-new-in-emacs-26-1}
\showURL{%
\tempurl}


\bibitem[Project(2021)]%
        {gir}
\bibfield{author}{\bibinfo{person}{The~GNOME Project}.} \bibinfo{year}{2021}\natexlab{}.
\newblock \bibinfo{title}{{GObject} {Introspection}}.
\newblock
\newblock
\urldef\tempurl%
\url{https://gi.readthedocs.io/en/latest/#gobject-introspection}
\showURL{%
\tempurl}


\bibitem[Rose et~al\mbox{.}(2003)]%
        {xml_rfc}
\bibfield{author}{\bibinfo{person}{Dr. Marshall~T. Rose}, \bibinfo{person}{Scott Hollenbeck}, {and} \bibinfo{person}{Larry~M Masinter}.} \bibinfo{year}{2003}\natexlab{}.
\newblock \bibinfo{title}{Guidelines for the use of extensible markup language ({XML}) within {IETF} protocols}.
\newblock
\newblock
\urldef\tempurl%
\url{https://doi.org/10.17487/RFC3470}
\showDOI{\tempurl}
\newblock
\shownote{Number: 3470 Series: Request for comments tex.howpublished: RFC 3470 tex.pagetotal: 28}.


\bibitem[Stallman et~al\mbox{.}(2004)]%
        {makefile}
\bibfield{author}{\bibinfo{person}{Richard Stallman}, \bibinfo{person}{Roland McGrath}, {and} \bibinfo{person}{Paul~D. Smith}.} \bibinfo{year}{2004}\natexlab{}.
\newblock \bibinfo{booktitle}{\emph{{GNU} {Make}: a program for directing recompliation ; {GNU} make version 3.81}}.
\newblock \bibinfo{publisher}{Free Software Foundation}, \bibinfo{address}{Boston, Mass}.
\newblock
\showISBNx{978-1-882114-83-2}


\bibitem[Team(2022)]%
        {gobject_boilerplate}
\bibfield{author}{\bibinfo{person}{The~GTK Team}.} \bibinfo{year}{2022}\natexlab{}.
\newblock \bibinfo{title}{How {To} define and implement a new {GObject}?}
\newblock
\newblock
\urldef\tempurl%
\url{http://tux.iar.unlp.edu.ar/~fede/manuales/gobject/howto-gobject.html}
\showURL{%
\tempurl}


\bibitem[{The GNOME Project}(2021)]%
        {vala}
\bibfield{author}{\bibinfo{person}{{The GNOME Project}}.} \bibinfo{year}{2021}\natexlab{}.
\newblock \bibinfo{title}{Vala - {Compiler} {Using} the {GObject} {Type} {System}}.
\newblock
\newblock
\urldef\tempurl%
\url{https://wiki.gnome.org/Projects/Vala}
\showURL{%
\tempurl}


\bibitem[Turcotte et~al\mbox{.}(2019)]%
        {pos_lua}
\bibfield{author}{\bibinfo{person}{Alexi Turcotte}, \bibinfo{person}{Ellen Arteca}, {and} \bibinfo{person}{Gregor Richards}.} \bibinfo{year}{2019}\natexlab{}.
\newblock \showarticletitle{Reasoning about foreign function interfaces without modelling the foreign language (artifact)}.
\newblock \bibinfo{journal}{\emph{Dagstuhl Artifacts Series}} \bibinfo{volume}{5}, \bibinfo{number}{2} (\bibinfo{year}{2019}), \bibinfo{pages}{9:1--9:2}.
\newblock
\showISSN{2509-8195}
\urldef\tempurl%
\url{https://doi.org/10.4230/DARTS.5.2.9}
\showDOI{\tempurl}
\newblock
\shownote{Place: Dagstuhl, Germany Publisher: Schloss Dagstuhl–Leibniz-Zentrum fuer Informatik}.


\bibitem[Wegiel and Krintz(2010)]%
        {colors}
\bibfield{author}{\bibinfo{person}{Michal Wegiel} {and} \bibinfo{person}{Chandra Krintz}.} \bibinfo{year}{2010}\natexlab{}.
\newblock \showarticletitle{Cross-language, type-safe, and transparent object sharing for co-located managed runtimes}.
\newblock \bibinfo{journal}{\emph{SIGPLAN Not.}} \bibinfo{volume}{45}, \bibinfo{number}{10} (\bibinfo{date}{Oct.} \bibinfo{year}{2010}), \bibinfo{pages}{223--240}.
\newblock
\showISSN{0362-1340}
\urldef\tempurl%
\url{https://doi.org/10.1145/1932682.1869479}
\showDOI{\tempurl}
\newblock
\shownote{Number of pages: 18 Place: New York, NY, USA Publisher: Association for Computing Machinery tex.issue\_date: October 2010}.


\bibitem[{Wikimedia Foundation}(2020)]%
        {wiki_ffi}
\bibfield{author}{\bibinfo{person}{{Wikimedia Foundation}}.} \bibinfo{year}{2020}\natexlab{}.
\newblock \bibinfo{title}{Foreign function interface}.
\newblock
\newblock
\urldef\tempurl%
\url{https://en.wikipedia.org/wiki/Foreign_function_interface}
\showURL{%
\tempurl}


\bibitem[{Wikipedia contributors}(2021)]%
        {jvm_hotspot}
\bibfield{author}{\bibinfo{person}{{Wikipedia contributors}}.} \bibinfo{year}{2021}\natexlab{}.
\newblock \bibinfo{title}{{HotSpot} (virtual machine) — {Wikipedia}, the free encyclopedia}.
\newblock
\newblock
\urldef\tempurl%
\url{https://en.wikipedia.org/w/index.php?title=HotSpot_(virtual_machine)&oldid=1045085695}
\showURL{%
\tempurl}


\bibitem[{Wikipedia contributors}(2022)]%
        {sql_stored_procedure}
\bibfield{author}{\bibinfo{person}{{Wikipedia contributors}}.} \bibinfo{year}{2022}\natexlab{}.
\newblock \bibinfo{title}{Stored procedure — {Wikipedia}, the free encyclopedia}.
\newblock
\newblock
\urldef\tempurl%
\url{https://en.wikipedia.org/w/index.php?title=Stored_procedure&oldid=1089982205}
\showURL{%
\tempurl}


\bibitem[{Wikipedia contributors}(2024)]%
        {name-mangling}
\bibfield{author}{\bibinfo{person}{{Wikipedia contributors}}.} \bibinfo{year}{2024}\natexlab{}.
\newblock \bibinfo{title}{Name mangling — {Wikipedia}, the free encyclopedia}.
\newblock
\newblock
\urldef\tempurl%
\url{https://en.wikipedia.org/w/index.php?title=Name_mangling&oldid=1193393826}
\showURL{%
\tempurl}


\bibitem[Wilmet(2019)]%
        {gobject}
\bibfield{author}{\bibinfo{person}{Sébastien Wilmet}.} \bibinfo{year}{2019}\natexlab{}.
\newblock \showarticletitle{The {GLib}/{GTK}+ development platform}.
\newblock \bibinfo{journal}{\emph{gnome.org}} (\bibinfo{year}{2019}).
\newblock


\bibitem[Yallop et~al\mbox{.}(2016)]%
        {generic_interop}
\bibfield{author}{\bibinfo{person}{Jeremy Yallop}, \bibinfo{person}{David Sheets}, {and} \bibinfo{person}{Anil Madhavapeddy}.} \bibinfo{year}{2016}\natexlab{}.
\newblock \showarticletitle{Declarative foreign function binding through generic programming}. In \bibinfo{booktitle}{\emph{{FLOPS}}}. \bibinfo{pages}{198--214}.
\newblock
\showISBNx{978-3-319-29603-6}
\urldef\tempurl%
\url{https://doi.org/10.1007/978-3-319-29604-3_13}
\showDOI{\tempurl}


\bibitem[Zhang(2023)]%
        {log4python}
\bibfield{author}{\bibinfo{person}{Xiang Zhang}.} \bibinfo{year}{2023}\natexlab{}.
\newblock \bibinfo{title}{log4python: log for python like java log4j2}.
\newblock
\newblock
\urldef\tempurl%
\url{https://pypi.org/project/log4python/}
\showURL{%
\tempurl}


\end{thebibliography}

\end{document}